\documentclass[reprint,nofootinbib,amsmath,amssymb,aps,prd]{revtex4-2}
\pdfoutput=1
\usepackage[T1]{fontenc}
\usepackage{amsmath}
\usepackage{siunitx}
\usepackage[colorlinks=true,citecolor=myurlcolor,linkcolor=myurlcolor,urlcolor=myurlcolor]{hyperref}
\usepackage{colortbl}\definecolor{myurlcolor}{rgb}{0,0,0.6}
\usepackage{color} 
\usepackage{graphicx}
\usepackage{dcolumn}
\usepackage{bm}
\usepackage{physics}
\DeclareMathAlphabet{\pazocal}{OMS}{zplm}{m}{n}
\usepackage{mathrsfs}
\usepackage{dsfont}
\usepackage{mathtools}
\graphicspath{{./Figures/}}

\newcommand{\kket}[1]{\left| \left. #1 \right> \right>}
\newcommand{\bbra}[1]{\left< \left<  #1 \right. \right|}
\newcommand{\brakket}[2]{\left<  #1 \right. \left|  \left. #2  \right> \right>}
\newcommand{\bbrakket}[2]{\left< \left<   #1 \right. \right.\left|  \left. #2  \right> \right>}
\newcommand{\eexpval}[1]{\left< \left<  #1 \right> \right>}

\begin{document}

\preprint{APS/123-QED}

\title{Robustness of the Page-Wootters construction across different pictures, states of the universe and system-clock interactions} 

\author{Simone Rijavec}
\email{simone.rijavec@physics.ox.ac.uk}
\affiliation{Clarendon Laboratory, University of Oxford, Parks Road, Oxford OX1 3PU, United Kingdom}

\date{\today}

\begin{abstract}
In quantum theory, the concept of time rests on shaky ground. One way to address this problem is to remove the usual background time parameter as a primitive entity and explain its emergence via correlations between physical systems. This approach was adopted by Page and Wootters (1983), who showed how time can emerge in a stationary quantum universe from the correlations between two of its subsystems, one of them acting as a clock for the other. In this work, I study the robustness of the Page-Wootters construction across different pictures, states of the universe and clock interactions, clarifying the role and the nature of the correlations between the subsystems of the universe. I start by showing how to formulate the Page-Wootters construction in the Heisenberg picture via a unitary change of basis. I consider both pure and mixed states of the universe and extend the analysis to include interactions between the clock and the other subsystem of the universe. The study reveals what kind of correlations are necessary for the construction to work. Interestingly, entanglement is not required as long as there are no interactions with the clock. The study also shows that these interactions can lead to a non-unitary evolution for some mixed states of the universe. In a simple two-level system, this aspect becomes relevant at scales where one would expect strong relativistic effects. At these scales, I also observe an inversion in the system's direction of time.
\end{abstract}

\maketitle

\section{Introduction.}
The concept of time in physics is related to a set of issues generally gathered under the umbrella term ``the problem of time'' \cite{barbour_end_2001,barbour_nature_2009,kiefer_does_2017,kiefer_concept_2009,isham_canonical_1993,kuchar_time_2011,anderson_problem_2017}. One of the main issues is that time usually appears in dynamical laws as a background (classical) parameter whose nature is completely detached from matter.  This is unsatisfactory, especially since matter and spacetime interact according to general relativity. The problem becomes even deeper in quantum theory, where one may argue that any system that can interact with a quantum system must itself be quantum (see, for example, \cite{marletto_quantum_2022}).

One way to address these issues is to discard time as a primitive entity external to physical systems and explain its emergence from the correlations between these systems.
One of the most prominent quantum versions of this ``timeless approach'' is the Page-Wootters (PW) construction \cite{page_evolution_1983}.
Page and Wootters showed how the usual time evolution of quantum theory can be recovered from the correlations between two subsystems of a stationary quantum universe. One of the subsystems is the system of interest, the other acts as a clock for the former.  In the PW construction, time is associated with an operator of the clock. The other system in the universe evolves relative to the eigenvalues of this time operator.

The PW construction is usually formulated in the Schr\"odinger picture for pure states of the universe and without interaction between the system and the clock. In this work, I study the robustness of the PW construction across different pictures, states of the universe and system-clock interactions. Studying each of these aspects is important on its own. First, although the Schr\"odinger picture (SP) and the Heisenberg picture (HP) are empirically equivalent, their description of physical reality is different, especially with regard to locality\footnote{The HP provides a local description of quantum theory \cite{deutsch_information_2000,deutsch_vindication_2012,raymond-robichaud_local-realistic_2021}. Since the universe in the PW construction is stationary, here locality refers to a local description of the system \textit{relative to the clock}.} \cite{deutsch_information_2000,deutsch_vindication_2012,raymond-robichaud_local-realistic_2021,marletto_interference_2021}. Mixed states of the universe, usually not considered in the literature, are interesting per se \cite{page_density_1986}. Finally, considering an interaction term between the system and the clock makes the construction more realistic \cite{smith_quantizing_2019}. 

Notably, the joint study of these aspects clarifies the role and nature of the system-clock correlations in the PW construction. These correlations are necessary to make the system evolve, but their nature is not fully clear in the literature. Most works trace it back to entanglement \cite{wootters_time_1984,moreva_time_2014,giovannetti_quantum_2015,marletto_evolution_2017}, while some argue that entanglement\footnote{Here and in the rest of this work, entanglement refers to the entanglement between system and clock in the \textit{kinematical} Hilbert space \cite{hohn_trinity_2021}.} is not necessary \cite{mendes_time_2019} or show that it is a feature linked to the SP \cite{loveridge_relative_2019,hohn_trinity_2021,ali_ahmad_quantum_2022}. This work reveals which correlations are necessary in different situations and how they transform in different pictures.

In the first part of this work, I start by reviewing the PW construction in the SP. 
The construction recovers the usual unitary time evolution of quantum theory for the system of interest after tracing out the clock's degrees of freedom. As a result, the HP can be easily recovered at the system level in the usual way \cite{page_evolution_1983}. However, this is different from moving to the HP at the level of the universe. In fact, the state vector of the universe is stationary, so the usual distinction between Schr\"odinger and the Heisenberg pictures is absent at this level, and the PW construction applied to the state of the universe will always lead to the SP description of the system.
Does this mean that the HP does not exist at the level of the universe?
Some works in the literature shed light on the problem through different approaches \cite{loveridge_relative_2019,hohn_switching_2019,chataignier_construction_2020,chataignier_relational_2021,hohn_how_2020,hohn_trinity_2021,hohn_equivalence_2021,kuypers_quantum_2022}. Here, I show that the PW construction can be formulated in the HP by performing a \textit{unitary} transformation on both the state vector and the observables of the universe, which transfers crucial system-clock correlations from the former to the latter. In this formulation, the observables of the system depend on the clock's time operator and the quantum nature of time is explicit. The usual time evolution of the system's observables can then be recovered via the relative-state construction in the HP  \cite{kuypers_everettian_2021}. 

The construction I outline here can be straightforwardly applied to mixed states of the universe. In addition to making the results of this work more general, the study of the HP formulation with mixed states of the universe sheds some light on what kind of system-clock correlations result in non-trivial time evolutions. Interestingly, system-clock entanglement is not a necessary property of the state of the universe, even in the SP.
Building on the results of \cite{kuypers_quantum_2022}, I also introduce a differential notation to express the results compactly and insightfully.

In the second part of this work, I consider interactions between the system and the clock, expanding some of the results of \cite{smith_quantizing_2019,singh_emergence_2023}. The objective is to demonstrate that the PW construction can be formulated in HP even in the presence of system-clock interactions and to investigate additional aspects of the construction with mixed states of the universe. I motivate the choice of a specific kind of interaction and show that this interaction can lead to a non-unitary evolution at the level of the system for some mixed states of the universe. Interestingly, the necessity of entanglement for a non-trivial system evolution is restored in the presence of this kind of system-clock interaction. This might have some implications for classical relational approaches to time. In this context, I also show that the SP and the HP do not exhaust all the possibilities. There are, in fact, an infinite number of pictures that differ on how the time dependence is split between the observables and the state of the universe. For example, it is possible to find an interaction picture where the observables of the system evolve freely while the state vector evolves under an interaction term.

In the last section, I apply the results of this work to some simple models. I show how to recover the HP for a two-level system and study the effect of system-clock interactions on the system's dynamics. I observe an inversion in the time direction of the system at scales where one would expect relativistic effects to be dominant. I also show that if the universe is in a mixed state, the system can lose its coherence in a characteristic time that depends on the strength of the system-clock interaction. 

\section{Review of the PW construction in the Schr\"odinger picture.} 
In this section, I review the assumptions of the PW construction and its SP formulation for both pure and mixed states of the universe.

\subsection{Pure states.}

The PW construction for pure states of the universe relies on the following assumptions \cite{page_evolution_1983,giovannetti_quantum_2015}:
\begin{enumerate}
    \item The universe can be divided into two subsystems. One is the system of interest $\mathfrak{S}$, the other acts as a clock $\mathfrak{C}$ for the first. The Hilbert space of the universe can be decomposed as $\mathscr{H}_{\mathfrak{U}}=\mathscr{H}_\mathfrak{S}\otimes \mathscr{H}_\mathfrak{C}$. In the ideal case, $\mathfrak{S}$ and $\mathfrak{C}$ are non-interacting, i.e. the Hamiltonian of the universe is given by:
            \begin{equation}
                \hat{\mathcal{H}}= \hat{H}\otimes \hat{\mathds{1}}_{\mathfrak{C}} + \hat{\mathds{1}}_{\mathfrak{S}} \otimes \hat{h}  ,
                \label{HU}
            \end{equation}
            where $\hat{H}$ is the Hamiltonian of the system and $\hat{h}$ is the Hamiltonian of the clock.
    
    \item The universe is in an eigenstate of its Hamiltonian $\hat{\mathcal{H}}$. Usually, the universe is assumed to be in the state with eigenvalue 0, but here I will consider any eigenstate $\kket{\Psi_{\mathcal{E}}}$: 
        \begin{equation}
           \hat{ \mathcal{H}}\kket{\Psi_{\mathcal{E}}} = \mathcal{E} \kket{\Psi_{\mathcal{E}}},
           \label{stationarity}
        \end{equation}
        where the double-ket notation is just a visual aid to remember that the states of the universe are defined on  $\mathscr{H}_{\mathfrak{S}}\otimes \mathscr{H}_{\mathfrak{C}}$.
        The states $\kket{\Psi_{\mathcal{E}}}$ satisfying the constraint of Eq. \eqref{stationarity} live in the physical Hilbert space $\mathscr{H}_{phy}$, a subspace of the Hilbert space of the universe $\mathscr{H}_{\mathfrak{U}}$ (also called \textit{kinematical} Hilbert space) \cite{smith_quantizing_2019}.

    \item The clock has an observable $\hat{t}$ conjugate to $\hat{h}$:
        \begin{equation}
           \left[\hat{t}, \hat{h}\right]= i,
            \label{th}
        \end{equation}
        where I have set $\hbar=1$.
        
\end{enumerate}

From these assumptions, the usual time evolution of the system can be derived in the following way \cite{giovannetti_quantum_2015}. Equation \eqref{th} implies that $e^{-i \hat{h} \theta}\ket{t}=\ket{t+\theta}\hspace{0.1cm}\forall\, \theta \in \mathds{R}$, where $\ket{t}$ is the eigenstate of $\hat{t}$ with eigenvalue $t$.
Using the identity decomposition in terms of $\{\ket{t}\}_t$, one can write
\begin{equation}
    \kket{\Psi_{\mathcal{E}}}=\int \text{d}t \ket{\psi_{\mathcal{E}}(t)}_{\mathfrak{S}}\ket{t}_{\mathfrak{C}} ,
    \label{initial_state}
\end{equation}
where it is assumed that the state of the system $\ket{\psi_{\mathcal{E}}(t)}_{\mathfrak{S}}$ is normalised at all times\footnote{This is guaranteed if one imposes, for example, that $\kket{\Psi_{\mathcal{E}}}$ must be normalized according to the scalar product $\bbrakket{\Phi}{\Psi}_{\text{phy}}\coloneqq\bbra{\Phi}\left(\ket{t}\bra{t}\otimes \mathds{1}_{\mathfrak{S}}\right)\kket{\Psi}$ which, due to Eqs. \eqref{HU} and \eqref{stationarity}, does not depend on $t$ \cite{smith_quantizing_2019}. }. 
Now, using equations \eqref{stationarity} and \eqref{HU}, one can find that
\begin{equation}
    e^{-i \mathcal{E} \theta} \kket{\Psi_{\mathcal{E}}}=e^{-i \hat{\mathcal{H}} \theta}  \kket{\Psi_{\mathcal{E}}} = \int \text{d}t\,  e^{-i \hat{H} \theta} \ket{\psi_{\mathcal{E}}(t)}_{\mathfrak{S}} \ket{t+\theta}_{\mathfrak{C}} ,
\end{equation}
and thus, equating this with Eq. \eqref{initial_state},
$\ket{\psi_{\mathcal{E}}(t)}_{\mathfrak{S}}=e^{-i\left(\hat{H}-\mathcal{E} \mathds{1}_{\mathfrak{S}} \right) t}\ket{\psi(0)}_{\mathfrak{S}} ,$ where I have fixed, without loss of generality, $\ket{\psi_{\mathcal{E}}(0)}=\ket{\psi(0)}\hspace{0.1cm} \forall\, \mathcal{E}$.
This means that the state of the system relative to the clock being in the state $\ket{t}$ is given by the partial inner product:
\begin{equation}
\prescript{}{\mathfrak{C}}{\brakket{t}{\Psi_{\mathcal{E}}}}=\ket{\psi_{\mathcal{E}}(t)}_{\mathfrak{S}} = e^{-i\left(\hat{H}-\mathcal{E} \mathds{1}_{\mathfrak{S}} \right)t}\ket{\psi(0)}_{\mathfrak{S}} ,
\label{relative_state_S}
\end{equation}
which shows that the relative-state $\ket{\psi(t)}_{\mathfrak{S}}$ satisfies the time-dependent Schr\"odinger equation with Hamiltonian $\hat{H}-\mathcal{E}$, where $\mathcal{E}$ is an unobservable shift in energy. In this construction, time emerges from the system-clock correlations: the time of the system is $t$ if the clock is in the state $\ket{t}_{\mathfrak{C}}$.

It is clear from Eq. \eqref{initial_state} that the PW construction leads to a non-trivial time evolution of the system only if there are at least two different $\ket{\psi_{\mathcal{E}}(t)}_{\mathfrak{S}}$ for two different values of $t$. This means that in the SP only \underline{entangled} pure states of the universe result in non-trivial time evolutions.

\subsection{Mixed states.}
The same construction outlined here can be applied to mixed states of the universe \cite{page_evolution_1983,marletto_evolution_2017}.
In this case, the stationarity condition of Eq. \eqref{stationarity} becomes \begin{equation}
    \left[\hat{\mathcal{R}}, \hat{\mathcal{H}}\right]=0,
    \label{stat_mixed}
\end{equation}
where $\hat{\mathcal{R}}$ is the density operator of the universe.
The state of the system relative to the clock being in the state $\ket{t}_\mathfrak{C}$ is now given by
\begin{equation}
    \hat{\rho}_{\mathfrak{S}}(t)= \frac{\underset{\mathfrak{C}}{\text{Tr}}\left[\hat{\mathcal{R}} \hat{\Pi}_t(\hat{t})\right]}
    {\text{Tr}\left[\hat{\mathcal{R}} \hat{\Pi}_t(\hat{t})\right] },
    \label{relative_rho_S}
\end{equation}
where $\hat{\Pi}_t(\hat{t})= \mathds{1}_{\mathfrak{S}} \otimes \ket{t}\bra{t}$ is the projector on the eigenspace of $\mathds{1}_{\mathfrak{S}} \otimes \hat{t}$ with eigenvalue $t$\footnote{The normalisation factor $\text{Tr}\left[\hat{\mathcal{R}} \hat{\Pi}_t(\hat{t})\right]$ does not depend on $t$ due to Eqs. \eqref{HU} and \eqref{stat_mixed}.}. The states $\hat{\mathcal{R}}$ such that $\text{Tr}\left[\hat{\mathcal{R}} \hat{\Pi}_t(\hat{t})\right]=0$ for at least one value of $t$ are excluded since they do not give rise to a physical state of the system\footnote{While the only stationary pure state of the universe such that $\brakket{t}{\Psi}=0$ for at least one value of $t$ is $\kket{\Psi}=0$, the class of stationary mixed states $\hat{\mathcal{R}}$ such that $\text{Tr}\left[\hat{\mathcal{R}} \hat{\Pi}_t(\hat{t})\right]=0$ is bigger.}.
Now, using Eqs. \eqref{HU}, \eqref{th} and \eqref{stat_mixed} one finds that

\begin{equation}
     \hat{\rho}_{\mathfrak{S}}(t)=e^{-i\hat{H}t} \hat{\rho}_{\mathfrak{S}}(0)  e^{i\hat{H}t}  \hspace{0.3cm} \forall \, t ,
\end{equation}
which is the usual Schr\"odinger-picture time evolution of a system with Hamiltonian $\hat{H}$. As I will discuss later, in the case of mixed states of the universe, system-clock entanglement is no longer necessary for a non-trivial time evolution of the system.

\section{PW construction in the Heisenberg picture.}\label{HP}
In this section, I show how to recover the PW construction in the HP at the level of the universe.
The approach and interpretation I outline here differ from
\cite{loveridge_relative_2019}, where the Heisenberg observables are recovered from other axioms and the choice of Heisenberg state is motivated differently. They also differ from \cite{chataignier_construction_2020, chataignier_relational_2021,hohn_trinity_2021,hohn_switching_2019,hohn_how_2020,hohn_equivalence_2021}, which generally start from invariant observables (i.e. observables that commute with a constraint operator), and from \cite{kuypers_quantum_2022}, which postulates that the HP is defined by observables of the system that obey a generalised Heisenberg equation\footnote{In Ref. \cite{kuypers_quantum_2022} the author requires that the Hamiltonian of the universe be equal to the Hamiltonian of the clock to prevent a real-valued time parameter from appearing in the equations of motion for the system's observables. Taken at face value, this requirement implies that the Hamiltonian of the clock must be identically zero. A correct interpretation of this statement, which is difficult to formulate in the framework of \cite{kuypers_quantum_2022}, is that the Hamiltonian of the universe in the HP must be equal to the Hamiltonian of the clock in the SP. This property finds a natural explanation in the approach described below.}.

Here, starting from the assumption that the universe is in a stationary state, I recover the HP description of the whole universe via a unitary transformation on both the observables and the state vector. The observables and Heisenberg state of \cite{loveridge_relative_2019} are recovered in a simple and well-motivated way; the assumptions of \cite{kuypers_quantum_2022} find a natural explanation; and the invariant observables appearing in \cite{loveridge_relative_2019,chataignier_construction_2020, chataignier_relational_2021,hohn_trinity_2021,hohn_switching_2019,hohn_how_2020,hohn_equivalence_2021} can be recovered via a group-average operation on the HP-observables of the system\footnote{Notice that the time measurements in \cite{loveridge_relative_2019,hohn_trinity_2021} are represented, in general, by a POVM while I restrict, for simplicity, to projective measurements. My results are in agreement with \cite{loveridge_relative_2019,hohn_trinity_2021} for this specific type of time measurements.}. 
Moreover, I extend the analysis to include mixed states of the universe.

\subsection{Pure states.}\label{HPpure}
For simplicity, let me focus only on the \textit{pure} state of the universe $\kket{\Psi_{0}}$ and omit the lower index. I will consider different eigenstates and mixed states later.
One can rewrite $\kket{\Psi}$ in Eq. \eqref{initial_state} in the following way \cite{giovannetti_quantum_2015}:
\begin{equation}
\kket{\Psi}=\int \text{d}t\, e^{-i\hat{H}t}\ket{\psi(0)}_{\mathfrak{S}} \ket{t}_{\mathfrak{C}}=e^{-i\hat{H}\otimes\hat{t}}\ket{\psi(0)}_{\mathfrak{S}}\ket{\text{TL}}_{\mathfrak{C}},
\label{StoH}
\end{equation}
where the ``Time Line'' state $\ket{\text{TL}}_{\mathfrak{C}}$ \cite{giovannetti_quantum_2015} is defined as $\ket{\text{TL}}_{\mathfrak{C}}\coloneqq \int \text{d}t\, \ket{t}_{\mathfrak{C}}=\sqrt{2\pi}\ket{h=0}_{\mathfrak{C}}$, with $\ket{h=0}_{\mathfrak{C}}$ the 0-eigenstate of $\hat{h}$.

Eq. \eqref{StoH} suggests that the transformation between the SP and the HP should be given by the unitary\footnote{Notice that this unitary is, apart from a phase and a constant, the same as the ``trivialisation map'' of \cite{hohn_trinity_2021}, and similar to a ``quantum reference frame transformation'' as in \cite{giacomini_quantum_2019}.} 
\begin{equation}
    \hat{\mathbb{U}}\coloneqq e^{-i\hat{H}\otimes\hat{t}} .
    \label{unitarySH}
\end{equation} 
For instance, if $\hat{O}$ is an observable of the system in the SP, the corresponding observable in the HP is: 
\begin{equation}
    \hat{O}^{(\mathbb{H})}= e^{i\hat{H}\otimes\hat{t}} \left(\hat{O}\otimes \mathds{1}_{\mathfrak{C}}\right) e^{-i\hat{H}\otimes\hat{t}} ,
    \label{qtime_op}
\end{equation}
which is an observable depending on the clock's time operator $\hat{t}$ and has support on $\mathscr{H}_{\mathfrak{U}}$.
The observables $\hat{H}$, $\hat{h}$ and $\hat{t}$ transform in the following way
\begin{align}
    &\hat{H}^{(\mathbb{H})}=\hat{\mathbb{U}}^{\dagger}\left( \hat{H} \otimes \mathds{1}_{\mathfrak{C}}\right) \hat{\mathbb{U}} = \hat{H}\otimes \mathds{1}_{\mathfrak{C}}, \label{newH} \\
    & \hat{h}^{(\mathbb{H})}=\hat{\mathbb{U}}^{\dagger} \left(\mathds{1}_{\mathfrak{S}} \otimes\hat{h} \right) \hat{\mathbb{U}} = \mathds{1}_{\mathfrak{S}} \otimes \hat{h}-\hat{H}\otimes \mathds{1}_{\mathfrak{C}} , \label{newh}\\
    & \hat{t}^{(\mathbb{H})}=\hat{\mathbb{U}}^{\dagger}\left(\mathds{1}_{\mathfrak{S}} \otimes \hat{t} \right)\hat{\mathbb{U}} = \mathds{1}_{\mathfrak{S}} \otimes\hat{t}, \label{newt}
\end{align}
which show that the Hamiltonian of the system and the time operator are invariant under $\hat{\mathbb{U}}$, while the Hamiltonian of the clock is not\footnote{Since the transformation between the SP and the HP is unitary, the relation $\left[\hat{h}^{(\mathbb{H})},\hat{O}^{(\mathbb{H})}\right]=\left[\hat{h}-\hat{H},\hat{O}^{(\mathbb{H})}\right]=0$ holds for every observable of the system $\hat{O}^{(\mathbb{H})}$. This commutation relation is the initial assumption of \cite{kuypers_quantum_2022}.}. However, the Hamiltonian of the whole system is not invariant:
\begin{equation}
    \hat{\mathcal{H}}^{(\mathbb{H})}=\hat{\mathbb{U}}^{\dagger} \hat{\mathcal{H}} \hat{\mathbb{U}} = \mathds{1}_{\mathfrak{S}} \otimes\hat{h}.
    \label{HTOT_H}
\end{equation} 
This is because here $\hat{\mathbb{U}}$ is a \textit{formal} transformation between the two pictures and not a time translation generated by $\hat{\mathcal{H}}$ (which usually defines the transformation between the two pictures).

For consistency with the SP, the Heisenberg state of the universe must be
\begin{equation}
    \kket{\Psi^{(\mathbb{H})}} \coloneqq \hat{\mathbb{U}}^{\dagger}\kket{\Psi} = \ket{\psi(0)}_{\mathfrak{S}}\ket{\text{TL}}_{\mathfrak{C}} ,
    \label{heisenberg_state}
\end{equation}
which is a separable state of clock and system.
The state $\kket{\Psi^{(\mathbb{H})}}$ still satisfies the stationarity constraint $\hat{\mathcal{H}}^{(\mathbb{H})}\kket{\Psi^{(\mathbb{H})}}=0$ but this is now entirely due to the state in the clock's Hilbert space: $\hat{h}\ket{\text{TL}}_{\mathfrak{C}}=0$.
Incidentally, this also shows that one is free to choose any initial state of the system $\ket{\psi(0)}_{\mathfrak{S}}$ while still preserving the constraint $\hat{\mathcal{H}}^{(\mathbb{H})}\kket{\Psi^{(\mathbb{H})}}=0$.

The Heisenberg operators $\hat{O}^{(\mathbb{H})}$ and the Heisenberg state  $\kket{\Psi^{(\mathbb{H})}}$ are all one needs to recover the time evolution of the system's observables and their expectation values.
Let me first notice that $\hat{O}^{(\mathbb{H})}$ contains the ``whole history'' of $\hat{O}$: $\eexpval{\hat{O}^{(\mathbb{H})}}=\int \text{d}t \, \prescript{}{\mathfrak{S}}{\bra{\psi(0)}}e^{i\hat{H}t}\hat{O} e^{-i\hat{H}t}\ket{\psi(0)}_{\mathfrak{S}}$,
where $\eexpval{\cdot}\coloneqq \bbra{\Psi^{(\mathbb{H})}}\cdot \kket{\Psi^{(\mathbb{H})}}$.
In order to get one time instance of $\hat{O}^{(\mathbb{H})}$, consider the following ``relative observables''\cite{kuypers_everettian_2021}:
\begin{equation}
     \hat{O}^{(\mathbb{H})}(t) \coloneqq \hat{O}^{(\mathbb{H})} \hat{\Pi}_t^{(\mathbb{H})}(\hat{t}) , 
      \label{rel_obs}
\end{equation}
where $\hat{\Pi}_t^{(\mathbb{H})}(\hat{t})\coloneqq\hat{\mathbb{U}}^{\dagger}\hat{\Pi}_t(\hat{t}) \hat{\mathbb{U}}=\hat{\Pi}_t(\hat{t}^{(\mathbb{H})})$. Eq. \eqref{newt} implies that $\hat{\Pi}_t^{(\mathbb{H})}(\hat{t})=\hat{\Pi}_t(\hat{t})$.
Eq. \eqref{rel_obs} describes the observables of the system relative to the clock being in the eigenstate of the time operator $\hat{t}$ with eigenvalue $t$. 

Since $\left[\hat{O}^{(\mathbb{H})}, \hat{\Pi}_t^{(\mathbb{H})}(\hat{t})\right]=0$ for every Heisenberg-picture observable $\hat{O}_H$ of the system, the relative observables form a ``relative sub-algebra'' at each time $t$ \cite{kuypers_everettian_2021}:
\begin{gather}
    \left( \hat{O}^{(\mathbb{H})} \hat{\Pi}_t^{(\mathbb{H})}(\hat{t})\right) ^2 =\left(\hat{O}^{(\mathbb{H})}\right)^2 \hat{\Pi}_t^{(\mathbb{H})}(\hat{t}), \\
    \left[\hat{O}^{(\mathbb{H})}\hat{\Pi}_t^{(\mathbb{H})}(\hat{t}),\hat{Q}^{(\mathbb{H})}\hat{\Pi}_t^{(\mathbb{H})}(\hat{t})\right]=\left[\hat{O}^{(\mathbb{H})},\hat{Q}^{(\mathbb{H})}\right]\hat{\Pi}_t^{(\mathbb{H})}(\hat{t})= \nonumber \\
    =\left[\hat{O},\hat{Q}\right]^{(\mathbb{H})}\hat{\Pi}_t^{(\mathbb{H})}(\hat{t}) ,
\end{gather}
for all the Heisenberg-picture observables $\hat{O}^{(\mathbb{H})}, \hat{Q}^{(\mathbb{H})}$ of the system, where $\left[\cdot,\cdot\right]^{(\mathbb{H})}\coloneqq \hat{\mathbb{U}}^{\dagger}\left[\cdot,\cdot\right]\hat{\mathbb{U}}$.
Furthermore, the expectation value of $\hat{O}^{(\mathbb{H})}$ relative to the clock being in the eigenstate with eigenvalue $t$ is given by \cite{kuypers_everettian_2021}
\begin{equation}
\frac{\eexpval{\hat{O}^{(\mathbb{H})}(t) }}{\eexpval{\hat{\Pi}_t^{(\mathbb{H})}(\hat{t})}}=\prescript{}{\mathfrak{S}}{\bra{\psi(0)}}e^{i\hat{H}t}\hat{O} e^{-i\hat{H}t}\ket{\psi(0)}_{\mathfrak{S}},
\end{equation}
in agreement with what one would have obtained from the PW construction in the SP at time $t$.

Since the Heisenberg state of the universe is separable with respect
to clock and system, one can find a ``reduced description'' of $\hat{O}_H(t)$ on the Hilbert space of the system given by 
\begin{equation}
    \left[\hat{O}^{(\mathbb{H})}(t)\right]_{\mathfrak{S}} \coloneqq \frac{\prescript{}{\mathfrak{C}}{\bra{\text{TL}}} \hat{O}^{(\mathbb{H})}(t)\ket{\text{TL}}_{\mathfrak{C}} }{\eexpval{
    \hat{\Pi}_t^{(\mathbb{H})}(\hat{t})}
    } =  e^{i\hat{H}t}\hat{O} e^{-i\hat{H}t}.
    \label{ctime_op}
\end{equation}
These are the usual Heisenberg-picture observables of a system with Hamiltonian $\hat{H}$.

%In the previous section, I remarked that only entangled (pure) states of the universe give rise to a non-trivial time evolution in the Schr\"odinger-picture. Indeed, if the state of the universe $\kket{\Psi}$ were separable the unitary $\hat{\mathbb{U}}$ such that $\hat{\mathbb{U}} \ket{\psi(0)}_{\mathfrak{S}}\ket{\text{TL}}_{\mathfrak{C}}=\kket{\Psi}$  would be of the form $\hat{V} \otimes \hat{W}$ with $\hat{V}$ and $\hat{W}$ local unitaries.  Consequently, all the HP observables of the system (Eq. \eqref{qtime_op}) would not be affected by $\hat{W}$, and thus they would not depend on the time operator $\hat{t}$ (i.e. they would be stationary).
Here, I have shown a possible way to recover the HP version of the PW construction starting from the specific form of the state of the universe. Is this construction unique? To answer this question, let me consider a generic unitary transformation $\hat{U}'$ between the two pictures. If $\kket{{\Psi'_{\mathcal{E}}}^{(\mathbb{H})}} \coloneqq \hat{U}'^{\dagger}\kket{\Psi_{\mathcal{E}}}$ is the associated Heisenberg state, I postulate that a \textit{good} transformation between the SP and the HP is defined by the following properties:
\begin{enumerate}
    \item $\hat{U}'^{\dagger}\left(\mathds{1}_\mathfrak{S}\otimes \hat{t}\right)\hat{U}' = \mathds{1}_\mathfrak{S}\otimes \hat{s}$, with $\hat{s}$ a self-adjoint an operator of the clock, so that the partial inner product and partial trace of Eqs. \eqref{relative_state_S} and \eqref{relative_rho_S} are still meaningful. This condition is satisfied in all the cases considered in this work.

    \item The transformation between the SP and the HP transfers all the ``encoded time evolution'' from the state of the universe to the observables. Specifically, $\kket{{\Psi'_{\mathcal{E}}}^{(\mathbb{H})}}$ is such that $\prescript{}{\mathfrak{C}}{\brakket{\Bar{s}'}{{\Psi'_{\mathcal{E}}}^{(\mathbb{H})}}} = e^{i\phi(s,s')} \prescript{}{\mathfrak{C}}{\brakket{\Bar{s}}{{\Psi'_{\mathcal{E}}}^{(\mathbb{H})}}} \hspace{0.1cm} \forall\, s,s'$, where $\ket{\Bar{s}}$ ($\ket{\Bar{s}'}$) is an eigenstate of the operator $\hat{s}$ defined above with eigenvalue $s$ ($s'$) and $\phi(s,s')$ is a phase.
\end{enumerate}
In Appendix \ref{propUpure}, I show that the possible unitary transformations $\hat{U}'$ between the SP and the HP differ from $\hat{\mathbb{U}}$ in Eq. \eqref{unitarySH} only by local unitary transformations\footnote{Local unitary means any unitary of the form $\hat{W}=\hat{W}_{\mathfrak{S}}\otimes\hat{W}_{\mathfrak{C}}$.}. These local unitaries do not affect the results above and thus lead to the Heisenberg picture evolution of the observables at the level of the system\footnote{In Appendix \ref{propUpure}, I also discuss a specific class of observables linked by a local unitary transformation of the clock. One can get ``gauge invariant'' observables  (i.e. commuting with $\hat{\mathcal{H}}$) of the system by a so-called ``group average'' operation over this class of observables.}. 

The unitary transformation $\hat{\mathbb{U}}$ represents the simplest choice of transformation between the SP and HP since it leaves the clock's time operator and the system's initial state invariant.
In this particular case, notice that $\kket{\Psi^{(\mathbb{H})}}$ is an eigenstate of $\hat{h}$ (with eigenvalue 0). Therefore, when one applies the Schr\"odinger-picture PW construction to $\kket{\Psi^{(\mathbb{H})}}$ one gets $\ket{\psi(t')}_{\mathfrak{S}}=\prescript{}{\mathfrak{C}}{\brakket{t'}{\Psi^{(\mathbb{H})}}}=\prescript{}{\mathfrak{C}}{\bra{t}e^{i\hat{h}(t'-t)}\kket{\Psi^{(\mathbb{H})}}}=\prescript{}{\mathfrak{C}}{\brakket{t}{\Psi^{(\mathbb{H})}}}=\ket{\psi(t)}_{\mathfrak{S}} \hspace{0.1cm} \forall t,t'$, that is, the state of the system  ``encoded'' in $\kket{\Psi^{(\mathbb{H})}}$ is stationary.

The results of Appendix \ref{propUpure} imply that the Heisenberg state of the universe must be \underline{separable} with respect to the system and  the clock. Does this mean that entanglement does not play a fundamental role in the construction? The answer is no. Entanglement is still essential in the construction. The SP-HP transformation simply moves the entanglement from the state of the universe to the observables. As always, entanglement depends not only on the state vector but also on the subalgebras of observables considered (which differ from the SP ones by the unitary transformation $\mathbb{U}$).

In Appendix \ref{propUpure}, I also show how my construction is related to the one of \cite{chataignier_construction_2020, chataignier_relational_2021,hohn_trinity_2021,hohn_switching_2019,hohn_how_2020,hohn_equivalence_2021}.

Finally, the construction carries over to different eigenstates $\kket{\Psi_{\mathcal{E}}}$ of $\hat{\mathcal{H}}$ with the same unitary $\hat{\mathbb{U}}$ but different Heisenberg states $\kket{\Psi_{\mathcal{E}}^{(\mathbb{H})}} =\ket{\psi(0)}\ket{\text{TL}_{\mathcal{E}}}$, where $\ket{\text{TL}_{\mathcal{E}}}\coloneqq \int \text{d}t\, e^{i\mathcal{E}t}\ket{t}$.

\subsection{Mixed states.}\label{HP_mixed}
Here, I derive the HP version of the PW construction starting from a mixed state of the universe. The construction could be applied, for instance, to any closed stationary sub-region of the universe entangled with the rest of the universe.  

For the density operator of the universe  $\hat{\mathcal{R}}$ in the SP, the stationarity constraint of Eq. \eqref{stat_mixed} implies that $\hat{\mathcal{R}}$ is of the form 
\begin{equation}
     \hat{\mathcal{R}} = \sum_{k,d_k} \nu_{k,d_k}\kket{{\mathcal{E}_k,d_k}}\bbra{{\mathcal{E}_k,d_k}}, 
     \label{initial_mixed}
 \end{equation}
where $\hat{\mathcal{H}}\kket{{\mathcal{E}_k,d_k}}=\mathcal{E}_k \kket{{\mathcal{E}_k,d_k}} \hspace{0.1cm} \forall \, k,d_k$, with $d_k$ the degeneracy label. For simplicity, I consider only a finite number of different energy eigenstates.
Using Eq. \eqref{StoH}, $\hat{\mathcal{R}}$ can be written as $\hat{\mathcal{R}}$ =$\hat{\mathbb{U}} \hat{\mathcal{R}}^{(\mathbb{H})} \hat{\mathbb{U}}^{\dagger}$ with $\hat{\mathbb{U}}$ the unitary defined in Eq. \eqref{unitarySH} and 
\begin{equation}
    \hat{\mathcal{R}}^{(\mathbb{H})}=
     \sum_{k,d_k} \nu_{k,d_k}\ket{\psi_{d_k}(0)}\bra{\psi_{d_k}(0)} \otimes \ket{\text{TL}_{\mathcal{E}_k}}\bra{\text{TL}_{\mathcal{E}_k}},
    \label{mixedHstate}
\end{equation}
where
$\braket{\psi_{d_k'}(0)}{\psi_{d_k}(0)}=\delta_{d_k d_k'}$. This state is separable. 
One can also write $\hat{\mathcal{R}}^{(\mathbb{H})}=\sum_{k} p_{k}\,\hat{\rho}_k(0) \otimes \ket{\text{TL}^{\mathcal{E}_k}}\bra{\text{TL}^{\mathcal{E}_k}}$ which shows that $\hat{\mathcal{R}}^{(\mathbb{H})}$ is a so-called \textit{one-way classically correlated} or \textit{quantum-classical} state \cite{horodecki_local_2005,piani_no-local-broadcasting_2008}. Due to Eqs. \eqref{stat_mixed} and \eqref{HTOT_H}, this state satisfies 
\begin{equation}
    \left[\hat{\mathcal{R}}^{(\mathbb{H})},\hat{h}\right]=0,
    \label{condition_H}
\end{equation} 
meaning that 
\begin{equation}
    \underset{\mathfrak{C}}{\text{Tr}}\left[\hat{\mathcal{R}}^{(\mathbb{H})} \hat{\Pi}_t(\hat{t})\right]= \underset{\mathfrak{C}}{\text{Tr}}\left[\hat{\mathcal{R}}^{(\mathbb{H})} \hat{\Pi}_{t'}(\hat{t})\right]
    \hspace{0.3cm} \forall \, t,t',
    \label{no_evo_mixed}
\end{equation}
that is, no time evolution is encoded in $\hat{\mathcal{R}}^{(\mathbb{H})}$.

The corresponding system's observables in the HP are defined as in Eq. \eqref{qtime_op}.
The relative observables $\hat{O}^{(\mathbb{H})}(t)$ are defined as in Eq. \eqref{rel_obs} and the expectation value of $\hat{O}^{(\mathbb{H})}$ relative to the clock being in the eigenstate with eigenvalue $t$ is given by
\begin{equation}
    \frac{{\text{Tr}}\left[ \hat{O}^{(\mathbb{H})}(t) \hat{\mathcal{R}}^{(\mathbb{H})}\right]}{{\text{Tr}} \left[\hat{\Pi}^{(\mathbb{H})}_t(\hat{t})\hat{\mathcal{R}}^{(\mathbb{H})} \right]} = \underset{\mathfrak{S}}{\text{Tr}}\left[e^{i\hat{H} t} \hat{O} e^{-i\hat{H} t} \hat{\rho}_{\mathfrak{S}}(0)\right],
    \label{exp_mix}
\end{equation}
where $\hat{\rho}_{\mathfrak{S}}(0)= \sum_k p_k \hat{\rho}_k(0)$.\\
Eq. \eqref{exp_mix} is what one would have obtained from the PW construction in the SP at time $t$.

Finally, since the Heisenberg state of Eq. \eqref{mixedHstate} is separable and since $\prescript{}{\mathfrak{C}}{\bra{\text{TL}_{\mathcal{E}_j}}}\hat{O}^{(\mathbb{H})}(t)\ket{\text{TL}_{\mathcal{E}_j}}_{\mathfrak{C}}=\prescript{}{\mathfrak{C}}{\bra{\text{TL}_{\mathcal{E}_k}}}\hat{O}^{(\mathbb{H})}(t)\ket{\text{TL}_{\mathcal{E}_k}}_{\mathfrak{C}}\hspace{0.1cm} \forall\, j,k$, one can define the reduced relative observables as
\begin{equation}
    \left[\hat{O}_H(t)\right]_{\mathfrak{S}} \coloneqq \frac{\underset{\mathfrak{C}}{\text{Tr}} \left[ \hat{O}^{(\mathbb{H})}(t)  \left(\mathds{1}_{\mathfrak{S}} \otimes\hat{\rho}_{\mathfrak{C}} \right) \right]}{{\text{Tr}} \left[\hat{\Pi}_t^{(\mathbb{H})}(\hat{t})\hat{\mathcal{R}} \right]} =  e^{i\hat{H}t}\hat{O}_S e^{-i\hat{H}t},
\end{equation}
 where $\hat{\rho}_{\mathfrak{C}}\coloneqq  \underset{\mathfrak{S}}{\text{Tr}} \left[ \hat{\mathcal{R}}^{(\mathbb{H})} \right] = \sum_k p_k \ket{\text{TL}_{\mathcal{E}_k}}\bra{\text{TL}_{\mathcal{E}_k}}_{\mathfrak{C}}$. These are the usual Heisenberg-evolved observables of the system.

In conclusion, the PW construction can be formulated in the HP also for mixed states of the universe. Furthermore, the mixed state of Eq. \eqref{initial_mixed} consists of an ensemble of eigenstates of $\hat{\mathcal{H}}$ with different eigenvalues and is thus more general than the pure state of Eq. \eqref{stationarity}.
If $\hat{\mathcal{R}}$ contains only states with the same eigenvalue $\mathcal{E}$, then the Heisenberg state of Eq. \eqref{mixedHstate}  becomes
\begin{equation}
    \hat{\mathcal{R}}^{(\mathbb{H})}=\hat{\rho}_{\mathfrak{S}}(0) \otimes \ket{\text{TL}_{\mathcal{E}}}\bra{\text{TL}_{\mathcal{E}}} ,  
\end{equation}
which is a product state. In general, however, the Heisenberg state, although separable, may have some other kinds of non-classical correlations \cite{modi_classical-quantum_2012}.

Is the SP-HP transformation outlined here unique? In Appendix \ref{propUmixed}, I show that---under the assumptions discussed at the end of Section \ref{HPpure}---all the other \textit{unitary} transformations between the two pictures differ from $\hat{\mathbb{U}}$ only by local unitaries. Similarly to the previous section, $\hat{\mathbb{U}}$ and $\hat{\mathcal{R}}^{(\mathbb{H})}$ represent the simplest choice of unitary and set of Heisenberg states\footnote{Notice that Eq. \eqref{mixedHstate} defines a \textit{set} of possible Heisenberg states. For example, $ \hat{\mathcal{R}}^{(\mathbb{H})}=\hat{\rho}_{\mathfrak{S}}(0) \otimes \ket{\text{TL}_{\mathcal{E}}}\bra{\text{TL}_{\mathcal{E}}}$ and $ \hat{\mathcal{R}}^{(\mathbb{H})}=\hat{\rho}_{\mathfrak{S}}(0) \otimes \mathds{1}_{\mathfrak{C}}$ are two such states that differ more than in just the choice of $\hat{\rho}_{\mathfrak{S}}(0)$ (the latter is recovered by letting $k$ become a continuous variable and integrating over all energy eigenvalues of $\hat{\mathcal{H}}$).}.  
As a result, the possible Heisenberg states of the universe commute with a local operator of the clock, similarly to Eq. \eqref{condition_H}.

This point shows another unique feature of the PW construction with mixed states of the universe. The set of states $\hat{\mathcal{R}}'$ such that $ \left[\hat{\mathcal{R}}',\hat{h}\right]=0$ is a subset of the set of all separable (mixed) states. This means that other \underline{separable} states can lead to a non-trivial system's evolution (in the SP). An example of such a state is 
\begin{equation}
    \hat{\mathcal{R}}'=\int \text{d}t \ket{\psi(t)}\bra{\psi(t)}\otimes \ket{t}\bra{t},
    \label{sep_mixed_state}
\end{equation}
with $\ket{\psi(t)}=e^{-i \hat{H} t}\ket{\psi(0)}$. This is a separable state that commutes with the Hamiltonian of the universe but gives rise to the usual time evolution at the level of the system\footnote{Notice that $\left[\hat{\mathcal{R}}',\hat{t}\right]=0$ but $\left[\hat{\mathcal{R}}',\hat{h}\right]\neq 0$. Also, moving to the HP one gets $\hat{\mathbb{U}}^{\dagger} \hat{\mathcal{R}}' \hat{\mathbb{U}}=\ket{\psi(0)}\bra{\psi(0)}\otimes \mathds{1}_{\mathfrak{C}}$.}.

Finally, if one has access only to the system, then it is not possible to distinguish an entangled state of the universe $\hat{\mathcal{R}}$ from the  state
\begin{equation}
    \hat{\mathcal{R}}_{\text{dec}} \coloneqq  \frac{1}{\mathcal{N}}\int \text{d}t\, \prescript{}{\mathfrak{C}}{\bra{t}}\hat{\mathcal{R}}\ket{t}_{\mathfrak{C}} \otimes \ket{t}\bra{t},
    \label{deco_mixed_state}
\end{equation}
which is a \textit{separable} stationary state of the universe with no coherences between the different eigenstates of $\hat{t}$\footnote{$\mathcal{N}\coloneqq \text{Tr}\left[\ket{t}\bra{t}\right]$ is an improper normalisation factor.}.

\section{Differential formulation.}\label{diff_form}
In this section, I introduce an alternative way of formulating the PW construction in terms of derivatives with respect to the time operator of the clock. Time-operator derivatives have already been mentioned in \cite{castro-ruiz_quantum_2020} and used in \cite{kuypers_quantum_2022} for the PW construction in the HP.
Here, I expand on these works by showing how the PW construction can be expressed in terms of time-operator derivatives both in the SP and the HP, and for both pure and mixed states of the universe. The result is a compact notation for the PW construction. I will later use this notation when introducing system-clock interactions.

Extending the notion of derivative to include variations over operators gives rise to some subtleties linked to the non-commutativity of the operators and the choice of the variation \cite{suzuki_quantum_1997}. In this work, I will focus on a specific type of operator derivatives, namely derivatives taken along the diagonal. This removes some of the subtleties mentioned above.
It is also important to notice that the results of this section hinge on the specific algebra of the clock\footnote{Specifically, Eq. \eqref{th} and the fact that $\hat{t}$ has a continuous, non-degenerate spectrum.} and the fact that the time operator is left unchanged when moving to the HP.

First, I introduce the derivative of a \textit{pure} state of the universe with respect to the time operator $\hat{t}$ of the clock:
\begin{equation}
    \frac{\text{d} \kket{\Psi}}{\text{d}\hat{t}} \coloneqq \int \text{d}t \,\frac{\text{d} \brakket{t}{\Psi}}{\text{d}t} \ket{t}_{\mathfrak{C}}  .
    \label{tder_pure}
\end{equation}
where $\text{d} \brakket{t}{\Psi}/\text{d}t=\lim_{\delta\rightarrow 0}\left[\brakket{t+\delta}{\Psi}-\brakket{t}{\Psi}\right]/\delta$ is the usual time derivative of a state vector. Eq. \eqref{tder_pure} resembles a directional derivative.
It is easy to show that 
\begin{equation}
    \frac{\text{d} \kket{\Psi}}{\text{d}\hat{t}} = i \left( \mathds{1}_{\mathfrak{S}} \otimes\hat{h} \right)\kket{\Psi}   ,
    \label{tder_pure2}
\end{equation}
which is similar to the action of momentum in single-particle quantum mechanics.
Using this, the stationarity condition of Eq. \eqref{stationarity} can be written as:
\begin{equation}
  i  \frac{\text{d} \kket{\Psi_{\mathcal{E}}}}{\text{d}\hat{t}} = \left(\hat{H}\otimes\mathds{1}_{\mathfrak{C}}-\mathcal{E}\right) \kket{\Psi_{\mathcal{E}}}  ,
  \label{q_Schr_eq}
\end{equation}
which has the form of the Schr\"odinger equation but with time-operator derivatives instead of ordinary time derivatives.
By taking the partial inner product of Eq. \eqref{q_Schr_eq} with $\ket{t}_{\mathfrak{C}}$, and using the definition of time-operator derivative of Eq. \eqref{tder_pure}, one gets, at any time $t$,
\begin{equation}
  i  \frac{\text{d} \ket{\psi_{\mathcal{E}}(t)}_{\mathfrak{S}}}{\text{d}t} = \left(\hat{H}-\mathcal{E}\right) \ket{\psi_{\mathcal{E}}(t)}_{\mathfrak{S}} ,
  \label{q_Schr_eq2}
\end{equation}
the usual Schr\"odinger equation on the system's Hilbert space (apart from an unobservable energy shift).

In order to move to the HP, I need to introduce time-operator derivatives of operators.
Let me define the derivative of an operator $\hat{A}$ (function of $\hat{t}$) with respect to $\hat{t}$ as \cite{suzuki_quantum_1997}
\begin{equation}
    \frac{\text{d} \hat{A}}{\text{d}\hat{t}} \coloneqq \lim_{\delta\rightarrow0} \frac{\hat{A}\left(\hat{t}+\delta \mathds{1}_{\mathfrak{C}}\right)-\hat{A}\left(\hat{t}\right)}{\delta}  ,
    \label{tder_op}
\end{equation}
where the variation has been taken along the diagonal. Since $\hat{t}$ and $\hat{h}$ have support only on the clock's Hilbert space, the time-operator derivative of $\hat{A}$ can also be written as
\begin{equation}
    \frac{\text{d} \hat{A}}{\text{d}\hat{t}} = \lim_{\delta\rightarrow0} \frac{\int\text{d}t\,\text{d}t'\,\ket{t}_{\mathfrak{C}}\bra{t+\delta}\hat{A}\ket{t'+\delta}_{\mathfrak{C}}\bra{t'}-\hat{A}}{\delta}  ,
    \label{tder_op2}
\end{equation}
which agrees with \cite{kuypers_quantum_2022} when $\hat{A}$ is a function of $\hat{t}$ and the system's operators only.
It is then easy to prove that
\begin{equation}
    \frac{\text{d} \hat{A}}{\text{d}\hat{t}} = i\left[\mathds{1}_{\mathfrak{S}} \otimes\hat{h},\hat{A}\right],
    \label{tder_op3}
\end{equation}
which also shows that a chain rule holds for the time-operator derivative.
 If one considers the operator $ \kket{\Psi}\bbra{\Psi}$, Eq. \eqref{tder_op2} implies that
 \begin{equation}
    \frac{\text{d} \kket{\Psi}\bbra{\Psi}}{\text{d}\hat{t}} = \frac{\text{d} \kket{\Psi}}{\text{d}\hat{t}}\bbra{\Psi}+\kket{\Psi}\frac{\text{d} \bbra{\Psi}}{\text{d}\hat{t}},
    \label{tder_density}
\end{equation}
with $\frac{\text{d} \kket{\Psi}}{\text{d}\hat{t}}$ as in Eq. \eqref{tder_pure}. This connects Eq. \eqref{tder_pure} and Eq. \eqref{tder_op} through a chain rule, which is why I use the same notation for both derivatives.

The notion of time-operator derivative of an operator and Eq. \eqref{tder_op3} allow me to recast the stationarity constraint of Eq. \eqref{stat_mixed} for a mixed state of the universe $\hat{\mathcal{R}}$ as
\begin{equation}
    \frac{\text{d} \hat{\mathcal{R}}}{\text{d}\hat{t}} = i\left[\hat{\mathcal{R}},\hat{H}\otimes\mathds{1}_{\mathfrak{C}}\right],
    \label{stat_mixed_diff}
\end{equation}
which is in the form of the Liouville–von Neumann equation but with time-operator derivatives instead of ordinary time derivatives.
Using the reduction procedure of Eq. \eqref{relative_rho_S} on Eq. \eqref{stat_mixed_diff} one gets
\begin{equation}
    \frac{\text{d} \hat{\rho}_{\mathfrak{S}}(t)}{\text{d}t} = i\left[\hat{\rho}_{\mathfrak{S}}(t),\hat{H}\right],
    \label{stat_mixed_diff2}
\end{equation}
the usual Liouville–von Neumann equation for the state of the system.

One can also use this notation for the PW construction in the HP. 
The time-operator derivative of $\hat{\mathbb{U}}$ in Eq. \eqref{unitarySH} is 
\begin{equation}
    \frac{\text{d} \hat{\mathbb{U}}}{\text{d}\hat{t}}=-i \left(\hat{H}\otimes\mathds{1}_{\mathfrak{C}}\right)\hat{\mathbb{U}}=-i\hat{H}^{(\mathbb{H})}\hat{\mathbb{U}},
    \label{tderHP1}
\end{equation}
and thus, using the chain rule on the HP observables of the system,
\begin{equation}
      \frac{\text{d} \hat{O}^{(\mathbb{H})}}{\text{d}\hat{t}}=i\left[\hat{H}\otimes\mathds{1}_{\mathfrak{C}},\hat{O}^{(\mathbb{H})}\right],
    \label{tderHP2}
\end{equation}
which is the usual equation of motion for the observables of the system in the HP but with time-operator derivatives instead of usual time derivatives. This is the equation found, in a different way, in \cite{kuypers_quantum_2022}.
The usual version of the equation at the level of the system's Hilbert space can be recovered via the procedure of Eq. \eqref{relative_rho_S}.

Notice that this result holds both for pure and mixed states of the universe. However, in the first case 
\begin{equation}
    \frac{\text{d} \kket{\Psi_{\mathcal{E}}^{(\mathbb{H})}}}{\text{d}\hat{t}} =\mathcal{E} \kket{\Psi_{\mathcal{E}}^{(\mathbb{H})}}  ,
    \label{tderHP3}
\end{equation}
while in the second
\begin{equation}
    \frac{\text{d} \hat{\mathcal{R}}^{(\mathbb{H})}}{\text{d}\hat{t}} =0 .
    \label{tderHP4}
\end{equation}
Both equations show that no time evolution is encoded in the Heisenberg state: $\hat{t}$ does not appear in $\hat{\mathcal{R}}^{(\mathbb{H})}$, while it appears only as a trivial phase in $\kket{\Psi_{\mathcal{E}}^{(\mathbb{H})}}$.

\section{System-clock interactions.}
\subsection{Restrictions on the type of interaction.}
The PW construction is usually formulated with no system-clock interactions. This idealisation significantly simplifies the construction but is not realistic. Furthermore, some works have studied interesting phenomena linked to the system-clock interaction, such as the emergence of a Newtonian potential or of a time dilation at the level of the system \cite{smith_quantizing_2019,smith_quantum_2020,singh_emergence_2023}.

In this section, I will consider a Hamiltonian of the universe of the following form:
\begin{equation}
    \hat{\mathcal{H}}= \hat{H}\otimes \hat{\mathds{1}}_{\mathfrak{C}} + \hat{\mathds{1}}_{\mathfrak{S}} \otimes \hat{h} + \hat{V},
    \label{HU_int}
\end{equation}
where $\hat{V}$ is the system-clock interaction. 
A detailed study of the effect of different types of interaction on the PW construction in the SP has been carried out in \cite{smith_quantizing_2019}.
Here, I will focus on a specific kind of interaction.

First, I will only consider interactions $\hat{V}$ such that
\begin{equation}
    \left[\hat{H}\otimes \hat{\mathds{1}}_{\mathfrak{C}} + \hat{\mathds{1}}_{\mathfrak{S}} \otimes \hat{h}, \hat{V}\right]=0.
    \label{en_cons}
\end{equation}
In ordinary (non-relativistic) quantum theory, this is one way to impose the conservation of the additive quantity $\hat{h}+\hat{H}$ \cite{marletto_quantum_2022}. To see this, notice that $e^{-i\hat{\mathcal{H}}t}\left(\hat{H}\otimes \hat{\mathds{1}}_{\mathfrak{C}} + \hat{\mathds{1}}_{\mathfrak{S}} \otimes \hat{h}\right)e^{i\hat{\mathcal{H}}t}=\hat{H}\otimes \hat{\mathds{1}}_{\mathfrak{C}} + \hat{\mathds{1}}_{\mathfrak{S}} \otimes \hat{h}$, that is, the quantity $\hat{H}\otimes \hat{\mathds{1}}_{\mathfrak{C}} + \hat{\mathds{1}}_{\mathfrak{S}} \otimes \hat{h}$ does not change in time. In the PW construction, the state of the universe is stationary, so even if Eq. \eqref{en_cons} did not hold, it would not be possible to observe a change in $\hat{h}+\hat{H}$. Nevertheless, here I take the position that conservation laws of additive quantities should hold as in Eq. \eqref{en_cons} even at the level of the universe despite the impossibility of observing a violation of them due to the constraints on the state of the universe.

The second constraint on $\hat{V}$ that I will impose is:
\begin{equation}
    \left[\mathds{1}_{\mathfrak{S}}\otimes\hat{h}, \hat{V}\right]=0.
    \label{time_ind_V}
\end{equation}
This is the same as requiring that $\text{d}\hat{V}/\text{d}\hat{t}=0$, that is, $\hat{t}$ must not appear in $\hat{V}$. As I will discuss later, this implies that the Hamiltonian that regulates the system's evolution must be \textit{time-independent}. Taken together with Eq. \eqref{en_cons}, Eq. \eqref{time_ind_V} implies that 
\begin{equation}
    \left[\hat{H}\otimes\mathds{1}_{\mathfrak{C}}, \hat{V}\right]=0,
    \label{add_constraint}
\end{equation}
and thus, since the spectrum of $\hat{h}$ is not degenerate, the eigenbasis of $\hat{V}$ must be of the form $\left\{\ket{E}\ket{\varepsilon}\right\}_{E,\varepsilon}$ where $\ket{E}_{\mathfrak{S}}$ ($\ket{\varepsilon}_{\mathfrak{C}}$) are the eigenstates of $\hat{H}$ ($\hat{h}$)\footnote{For simplicity, and without loss of generality, I am not considering degeneracies in the spectrum of $\hat{H}$.}.
 
The last constraint that I will consider is that $\hat{V}$ must not contain powers of $\hat{h}$ higher than 1. In other terms,
\begin{equation}
    \frac{\text{d}^k\hat{V}}{\text{d}\hat{h}^k}=0 \hspace{0.3cm}\forall\, k\geq 2,
    \label{1order}
\end{equation} 
where higher-order derivatives with respect to an operator are defined by a recursive application of the operator derivatives defined in Sec. \ref{diff_form}.
As I will discuss later, this condition ensures that only first-order time derivatives appear in the system's equations of motion. 

Given these constraints, all the allowed system-clock interactions are of the form:
\begin{equation}
    \hat{V}=\hat{X}\otimes\hat{h},
    \label{allowed_V}
\end{equation}
where $\hat{X}$ is a self-adjoint operator of the system such that $\left[\hat{X},\hat{H}\right]=0$. If the only information we know about the systems is its Hamiltonian, we can only consider interactions where $\hat{X}=f(\hat{H})$:
\begin{equation}
    \hat{V}=f(\hat{H})\otimes\hat{h}.
    \label{allowed_V2}
\end{equation}
The most instructive example is the one considered in \cite{castro_ruiz_entanglement_2017,smith_quantizing_2019,castro-ruiz_quantum_2020,singh_emergence_2023}:
\begin{equation}
    \hat{V}^G=\frac{1}{\Lambda}\hat{H}\otimes\hat{h},
    \label{V_G}
\end{equation}
where the label $G$ stands for ``gravitational'' due to its connection with the gravitational potential with some post-Newtonian corrections as in \cite{castro_ruiz_entanglement_2017}. Some works in the literature \cite{smith_quantizing_2019,smith_quantum_2020,singh_emergence_2023} have shown that this interaction leads to some gravitational-like effects on the system when choosing $\Lambda=dc^4/G$, with $d$ a characteristic length of the system.

\subsection{Explicit solutions for pure states.}\label{int_pure}
Given a system-clock interaction as in Eq. \eqref{allowed_V2}, I show how to derive an explicit solution in the case of a pure state of the universe. A similar study was already performed in \cite{smith_quantizing_2019,singh_emergence_2023} for pure states of the universe. Here, I generalise the analysis by considering different eigenvalues of $\hat{\mathcal{H}}$, which result in different dynamics at the level of the system. This effect becomes particularly relevant for mixed states of the universe (considered in the next subsection). I also show how to recover the HP. The approach followed here is in the same spirit of \cite{singh_emergence_2023}.

First, let me notice that the spectrum of $\hat{V}$ is $\left\{f(E)\varepsilon\right\}_{E,\varepsilon}$ with $E$ and $\varepsilon$ eigenvalues of $\hat{H}$ and $\hat{h}$, respectively.
Since the state of the universe $\kket{\Psi_0}$ can be written, in general, as\footnote{Here, I am assuming that the spectrum of $\hat{H}$ is continuous.}  
\begin{equation}
    \kket{\Psi_0}=\int \text{d}E\,\text{d}\varepsilon\, \lambda_{E\varepsilon}\ket{E}\ket{\varepsilon},
    \label{pure_int_state}
\end{equation}
the stationarity condition $\mathcal{H}\kket{\Psi_0}=0$ reads
\begin{equation}
    \hat{\mathcal{H}}\kket{\Psi_0}=\int \text{d}E\,\text{d}\varepsilon\, \left(E+\varepsilon+f(E)\varepsilon\right)\lambda_{E\varepsilon}\ket{E}\ket{\varepsilon}=0,
\end{equation}
and thus, for a given $E$ and $\varepsilon$, either $\lambda_{E\varepsilon}= 0$ or
\begin{equation}
    \varepsilon=g(E)\coloneqq -\frac{E}{1+f(E)} ,
    \label{gE}
\end{equation}
which is well defined for all $E$ such that $f(E)\neq-1$\footnote{If $f(E^*)=-1$ for some $E^*$ then there is no solution. In this case $\lambda_{E^*\varepsilon}$ must be zero for any $\varepsilon$. Usually, the interaction considered is weak enough so that $f(E)=-1$ only for very large negative energies. In Sec. \ref{models}, I will consider the effects of a strong interaction on a simple model.}.
Therefore, Eq. \eqref{pure_int_state} becomes
\begin{align}
    \kket{\Psi_0}=&\int \text{d}E\, \lambda_{Eg(E)}\ket{E}\ket{g(E)}= \nonumber \\
    = &\int \text{d}t\,\left(\int\text{d}E\, \lambda_{Eg(E)}\frac{e^{i g(E) t}}{\sqrt{2\pi}}\ket{E}\right) \ket{t}= \nonumber \\
   =& \int \text{d}t \,e^{ig(\hat{H})t}\ket{\psi_0(0)}\ket{t} ,
   \label{pure_int_state2}
\end{align}
where $\ket{\psi_0(0)}\coloneqq\int\text{d}E\, \frac{\lambda_{Eg(E)}}{\sqrt{2\pi}}\ket{E}$ with $\lambda_{Eg(E)}=0$ wherever $g(E)$ is not defined\footnote{Notice that the normalisation of $\ket{\psi(t)}$ is guaranteed at all times if $\bra{\psi(0)}\ket{\psi(0)}=1$.}.
So, by taking the partial inner product with $\prescript{}{\mathfrak{C}}{\bra{t}}$, one recovers a unitary evolution of the system in the SP with an effective Hamiltonian
\begin{equation}
    \hat{H}_{eff}\coloneqq - g(\hat{H}) .
    \label{H_eff}
\end{equation}
For instance, in the case of the potential $\hat{V}^G$ of Eq. \eqref{V_G} one has \cite{singh_emergence_2023}
\begin{equation}
    \hat{H}_{eff}^G=\frac{\hat{H}}{1+\frac{\hat{H}}{\Lambda}}.
    \label{H_eff_G}
\end{equation}
Importantly, the Hamiltonians of Eqs. \eqref{H_eff} and \eqref{H_eff_G} are \underline{self-adjoint}. This property is guaranteed by Eq. \eqref{add_constraint}. If Eq. \eqref{add_constraint} does not hold, the effective Hamiltonian will be, in general, non-self-adjoint \cite{paiva_non-inertial_2022}.

From Eq. \eqref{pure_int_state2} it is also easy to see how to move to the HP, since $\kket{\Psi_0}=\hat{\mathds{U}}'\ket{\psi(0)}\ket{\text{TL}}$ where the unitary
\begin{equation}
    \hat{\mathds{U}}'\coloneqq e^{-i\hat{H}_{eff}\otimes\hat{t}},
    \label{U_int}
\end{equation}
can be used to move between the SP and the HP as in Sec. \ref{HPpure}.
Under $\hat{\mathds{U}}'$ Eqs. \eqref{newH}-\eqref{newt} change to
\begin{align}
    &\hat{H}^{(\mathbb{H})}= \hat{H}\otimes \mathds{1}_{\mathfrak{C}}, \label{newHint} \\
    & \hat{h}^{(\mathbb{H})}=\mathds{1}_{\mathfrak{S}} \otimes \hat{h}-\hat{H}_{eff}\otimes \mathds{1}_{\mathfrak{C}} , \label{newhint}\\
    & \hat{t}^{(\mathbb{H})}= \mathds{1}_{\mathfrak{S}} \otimes\hat{t},\label{newtint}
\end{align}
crucially leaving $\hat{t}$ invariant so that $\hat{\mathds{U}}'$ and $\ket{\psi(0)}\ket{\text{TL}}$ represent a good SP-HP transformation and Heisenberg state according to the requirements of Section \ref{HPpure}.
Thus, the HP PW construction developed in Sec. \ref{HPpure} can be straightforwardly applied to this case by replacing $\hat{H}$ with $\hat{H}_{eff}$.

Also notice that $\hat{H}_{eff}$ and $\hat{V}$ transform in the following way
\begin{align}
    &\hat{H}_{eff}^{(\mathbb{H})}= \hat{H}_{eff}\otimes \mathds{1}_{\mathfrak{C}}, \label{newHeffint} \\
    & \hat{V}^{(\mathbb{H})}= \hat{V}+i\hat{H}_{eff} \cdot \left[\hat{t},\hat{V}\right]=\hat{V}-\hat{H}_{eff} \cdot \frac{\text{d}\hat{V}}{\text{d}\hat{h}}, \label{newVint}
\end{align}
where Eq. \eqref{newVint} follows from Eqs. \eqref{add_constraint} and \eqref{1order}.
Interestingly, when using the gravitational interaction of Eq. \eqref{V_G} the Hamiltonian of the universe in the HP becomes $\hat{\mathcal{H}}^{(\mathbb{H})}=\mathds{1}_{\mathfrak{S}} \otimes \hat{h}+\hat{V}^G$.

An important difference between the interacting and non-interacting PW constructions emerges when considering different eigenstates of $\hat{\mathcal{H}}$. For an eigenstate of $\hat{\mathcal{H}}$ with eigenvalue $\mathcal{E}$, the function $g(E)$ in Eq. \eqref{gE} becomes:
\begin{equation}
   g_{\mathcal{E}}(E)\coloneqq - \frac{E-\mathcal{E}}{1+f(E)} ,
    \label{gE2}
\end{equation}
and thus the effective Hamiltonian of the system changes to $\hat{H}_{eff}(\mathcal{E})\coloneqq - g_{\mathcal{E}}(\hat{H})$.
In the case of the interaction $\hat{V}^G$, the effective Hamiltonian is given by
\begin{equation}
    \hat{H}_{eff}^{G}(\mathcal{E})=\frac{\hat{H}-\mathcal{E}}{1+\frac{\hat{H}}{\Lambda}}\approx \hat{H}_{eff}^{G}(0) - \mathcal{E} + \frac{\mathcal{E}\hat{H}}{\Lambda},
    \label{H_eff_G_eps}
\end{equation}
with the approximation holding as long as $\Lambda$ is much bigger than the typical energy of the system.
This leads to a non-trivial modification in the dynamics of the system. The difference in the system's dynamics between two eigenstates of the universe with eigenvalues $\mathcal{E}$ and $\mathcal{E}'$ can be quantified in the following way:
\begin{equation}
    \abs{\braket{\psi^\mathcal{E'}(t)}{\psi^\mathcal{E}(t)}} \approx \abs{\matrixel{\psi(0)}{e^{i\left(\mathcal{E'}-\mathcal{E}\right) \hat{H} t/\Lambda}} {\psi(0)}},
\end{equation}
where I have assumed that the initial of the system is the same in both cases.

\subsection{Explicit solutions for mixed states.}
Here, I  assume that the state of the universe is a mixed state $\hat{\mathcal{R}}$.
The general form of $\hat{\mathcal{R}}$ is given by Eq. \eqref{initial_mixed} which, using Eq. \eqref{pure_int_state2}, can also be written as\footnote{Here, I am also assuming that the spectrum of $\hat{H}$ is continuous. For simplicity, I am taking the sum over a discrete number of eigenstates of $\hat{\mathcal{H}}$.}  
\begin{equation}
     \hat{\mathcal{R}} = \int\text{d}t\text{d}t'\sum_{k}p_k e^{ig_k(\hat{H})t} \hat{\rho}_k(0) \otimes\ket{t}\bra{t'}e^{-ig_k(\hat{H})t'},
     \label{initial_mixed_int} 
 \end{equation}
where $\hat{\rho}_k(0)\coloneqq \frac{1}{p_k}\sum_{d_k}\nu_{k,d_k}\int\text{d}E\text{d}E'\,\frac{\lambda_{E}^{d_k}\lambda_{E'}^{d_k *}}{2\pi}\ket{E}\bra{E'}$ with $p_k$ such that $\text{Tr}\left[\hat{\rho}_k(0)\right]=1$, and $g_k(\hat{H})\coloneqq g_{\mathcal{E}_k}(\hat{H})$.
Thus, using Eqs. \eqref{relative_rho_S} and \eqref{H_eff}, the state of the system at time $t$ reads
 \begin{equation}
    \hat{\rho}_{\mathfrak{S}}(t) = \sum_{k}p_k e^{-i\hat{H}_{eff}(\mathcal{E}_k)t} \hat{\rho}_k(0) e^{i\hat{H}_{eff}(\mathcal{E}_k)t},
     \label{relative_mixed_int}
\end{equation}
where the coefficients $p_k$ are greater than 0 and such that $\sum_k p_k=1$.
Importantly, if the state of the universe consists of eigenstates of $\hat{\mathcal{H}}$ with different eigenvalues, the evolution of the system is no longer unitary overall\footnote{Importantly, the map $\hat{\rho}_{\mathfrak{S}}(0)\rightarrow\hat{\rho}_{\mathfrak{S}}(t)$ preserves the positivity and the trace of $\hat{\rho}_{\mathfrak{S}}(0)$.}. Instead, states of the system associated with different eigensectors of $\hat{\mathcal{H}}$ evolve with Hamiltonians which, due to Eq. \eqref{gE2}, differ by more than just a trivial energy shift. In the special case where all the $\hat{\rho}_k(0)$ are equal, the evolution of $\hat{\rho}_{\mathfrak{S}}$ is given by the CPTP map $\hat{\rho}_{\mathfrak{S}}(t)=\sum_k \hat{B}_k(t)\hat{\rho}_{\mathfrak{S}}(0)\hat{B}_k^{\dagger}(t)$ with Kraus operators $\hat{B}_k(t)\coloneqq \sqrt{p_k} e^{-i\hat{H}_{eff}(\mathcal{E}_k)t}$.

The evolution of the system can be unitary only if the state of the universe belongs to a single eigensector of $\hat{\mathcal{H}}$\footnote{For a different kind of non-unitarity due to a non-ideal clock see \cite{mendes_non-linear_2021}.}.
An interesting consequence is that only \underline{entangled} states of the universe give rise to a non-trivial unitary time evolution on the system's Hilbert space\footnote{This can be seen by restricting Eq. \eqref{initial_mixed_int} to only one term of the sum.}. This means that separable states such as those in Eqs. \eqref{sep_mixed_state}-\eqref{deco_mixed_state} are no longer allowed. 

Finally, in order to move to the HP, let me rewrite Eq. \eqref{initial_mixed_int} as
\begin{equation}
     \hat{\mathcal{R}} = \sum_{k}p_k e^{-i\hat{H}_{eff}(\mathcal{E}_k)\otimes\hat{t}} \hat{\rho}_k(0) \otimes\ket{\text{TL}}\bra{\text{TL}}e^{i\hat{H}_{eff}(\mathcal{E}_k)\otimes\hat{t}}.
     \label{initial_mixed_intH} 
 \end{equation}
If there is more than one term in the sum, then there is no unitary to move between the SP and the HP. In the special case where all the $\hat{\rho}_k(0)$ are equal, we can write 
\begin{equation}
    \hat{\mathcal{R}}=\sum_k \hat{\mathds{B}}_k\hat{\mathcal{R}}^{(\mathbb{H})}\hat{\mathds{B}}_k^{\dagger},
     \label{initial_mixed_intH2} 
\end{equation}
with $\hat{\mathcal{R}}^{(\mathbb{H})}\coloneqq\hat{\rho}_{\mathfrak{S}}(0)\otimes\ket{\text{TL}}\bra{\text{TL}}$ and Kraus operators $\hat{\mathds{B}}_k\coloneqq \sqrt{p_k} e^{-i\hat{H}_{eff}(\mathcal{E}_k)\otimes\hat{t}}$. To recover the HP, we can act with the adjoint map on the observables of the universe and use the Heisenberg state $\hat{\mathcal{R}}^{(\mathbb{H})}$.
For instance, the observables of the system in HP are given by
\begin{equation}
    \hat{O}^{(\mathbb{H})}= \sum_k \hat{\mathds{B}}_k^{\dagger}\left(\hat{O}\otimes \mathds{1}_{\mathfrak{C}}\right) \hat{\mathds{B}}_k .
    \label{qtime_op_mixed_int}
\end{equation}

The construction then follows the lines of sections \ref{HP_mixed} and \ref{int_pure}.
Notice that the adjoint map leaves the operator $\hat{t}$ unchanged, and the Heisenberg state is such that $\left[\hat{\mathcal{R}}^{(\mathbb{H})},\hat{h}\right]=0$, in agreement with Eq. \eqref{condition_H}.

\subsection{Differential formulation and interaction picture.}
The results above can be represented compactly using the time operator derivatives of Sec. \ref{diff_form}.
For pure states of the universe and the interaction $\hat{V}^G$, the stationarity condition can be written as:
\begin{equation}
    i\left(\mathds{1}_{\mathfrak{U}}+\frac{\hat{H}}{\Lambda}\otimes\mathds{1}_{\mathfrak{C}}\right)  \frac{\text{d} \kket{\Psi_{\mathcal{E}}}}{\text{d}\hat{t}} = \left(\hat{H}\otimes\mathds{1}_{\mathfrak{C}}-\mathcal{E}\right) \kket{\Psi_{\mathcal{E}}} , 
\end{equation}
and thus 
\begin{equation}
    i  \frac{\text{d} \kket{\Psi_{\mathcal{E}}}}{\text{d}\hat{t}} =\left( \hat{H}_{eff}^{G}(\mathcal{E})\otimes\mathds{1}_{\mathfrak{C}}\right) \kket{\Psi_{\mathcal{E}}},
    \label{diff_int}
\end{equation}
with $\hat{H}_{eff}^{G}(\mathcal{E})$ the effective Hamiltonian of Eq. \eqref{H_eff_G_eps}. Interestingly, it is not possible to recover an analogous operatorial Liouville-von Neumann equation for mixed states of the universe. This is linked to the fact that the system's evolution is no longer unitary for a general mixed state.

The differential notation is also helpful for quickly grasping some aspects of the PW construction with system-clock interaction. For instance, if the interaction contains powers of $\hat{h}$ greater than 1, then the Schr\"odinger equation will have higher-order time derivatives. Alternatively, if $\hat{V}$ does not commute with $\hat{h}$ --- that is, $\hat{t}$ appears in $\hat{V}$ --- then the effective Hamiltonian of the system will be time-dependent \cite{smith_quantizing_2019}.

Finally, let me notice that the SP and the HP do not exhaust all the possibilities. There can be infinite pictures that differ on how the time dependence is split between the observables and the state of the universe. For example, there is an interaction picture where the observables are evolved with the ``free'' unitary $\hat{\mathbb{U}}$ of Eq. \eqref{unitarySH}, while the state of the universe $\kket{\Psi^{(\mathbb{I})}}\coloneqq\hat{\mathbb{U}}^{\dagger}\kket{\Psi}$ obeys the following equation:
\begin{equation}
     i\frac{\text{d} \kket{\Psi^{(\mathbb{I})}}}{\text{d}\hat{t}} = \hat{V}^{(\mathbb{I})} \kket{\Psi^{(\mathbb{I})}},
     \label{int_picture}
\end{equation}
where $\hat{V}^{(\mathbb{I})}\coloneqq \hat{\mathbb{U}}^{\dagger}\hat{V}\hat{\mathbb{U}}$ is the system-clock interaction in the new picture. This shows that the state is still dependent on $\hat{t}$ and thus not all the time dependence has been transferred to the observables. In particular, $\kket{\Psi^{(\mathbb{I})}}$ is still entangled.

\section{Some simple models.}\label{models}
Here I show how the results obtained above can be applied to some simple scenarios.
I start by showing how the PW construction for some simple qubit models can be formulated in the HP. Then, I consider the effects of a system clock interaction between a qubit and an ``ideal clock''\footnote{Here, ``ideal clock'' means a clock with an algebra as in Eq. \eqref{th}.}. Some unique effects arise when system and clock are strongly interacting. I also study these effects in a massive two-level system. Finally, I consider how the system's dynamics changes when the universe is in a mixed state.

\subsection{Qubit and ideal clock.}
Let me consider a model where the system of interest is a qubit and the clock is an ideal clock with two conjugate observables $\hat{t}$ and $\hat{h}$. Let me assume that the Hamiltonian of system+clock is $\hat{\mathcal{H}}=\hat{H}\otimes\mathds{1}_{\mathfrak{C}}+\mathds{1}_{\mathfrak{S}}\otimes\hat{h}$ with $\hat{H}=\sigma_x/2$. Here $\sigma_{x,y,z}$ denote the Pauli matrices. Let me also assume that the state of system+clock is an eigenstate of $\hat{\mathcal{H}}$ with eigenvalue 0 and that the system is in the state $\ket{\psi(0)}_{\mathfrak{S}}$ at time $t=0$. Since all the qubit's observables can be expressed in terms of two generators of its algebra, one can choose to keep only track of $\left(\sigma_x\otimes\mathds{1}_{\mathfrak{C}}, \sigma_z\otimes\mathds{1}_{\mathfrak{C}}\right)$ \cite{gottesman_heisenberg_1998}. 

To move to the HP, one can choose the Heisenberg state $\ket{\psi(0)}_{\mathfrak{S}}\ket{\text{TL}}_{\mathfrak{C}}$ and act with the unitary $\hat{\mathbb{U}}=e^{-i\hat{H}\otimes\hat{t}}$ on the generators obtaining
\begin{align}
&\left(\sigma_x\otimes\mathds{1}_{\mathfrak{C}},\sigma_z\otimes\mathds{1}_{\mathfrak{C}}\right)\xrightarrow{}\\
   & \left(\sigma_x\otimes\mathds{1}_{\mathfrak{C}},\sigma_z \otimes\text{cos}\left(\hat{t}\right)+\sigma_y\otimes\text{sin}\left(\hat{t}\right)\right), \nonumber 
\end{align}
where the dependence of the system's observables on the time operator $\hat{t}$ is explicit. Moreover, the qubit's generators relative to the clock being in the eigenstate with eigenvalue $t$ are given by $\left(\sigma_x \hat{\Pi}_t(\hat{t}),\left[\sigma_z \,\text{cos}\left(t\right)+\sigma_y\,\text{sin}\left(t\right)\right]\hat{\Pi}_t(\hat{t})\right)$. Using equation \eqref{ctime_op} one can get the evolved generators of the system $\left(\sigma_x,\sigma_z \,\text{cos}\left(t\right)+\sigma_y\,\text{sin}\left(t\right)\right)$ that one would have obtained through the Heisenberg equation.

\subsection{Qubit system and clock.}
As a further simplification, let me assume that the clock consists of a single qubit \cite{moreva_time_2014} and that the time operator $\hat{t}$ is given by $\hat{t}\coloneqq t_0\ket{0}\bra{0}_{\mathfrak{C}} + t_1 \ket{1}\bra{1}_{\mathfrak{C}}$ where $\ket{0}_{\mathfrak{C}}$ ($\ket{1}_{\mathfrak{C}}$) is the +1 (-1) eigenstate of $\hat{\mathds{1}}_{\mathfrak{S}}\otimes \sigma_z$.
In this case, to avoid a trivial evolution, one must consider only non-degenerate Hamiltonians of clock and system such that their sum has a doubly degenerate zero eigenvalue \footnote{In the previous example this was guaranteed for any non-degenerate Hamiltonian of the qubit since the spectrum of the clock Hamiltonian was $\mathds{R}$.}. For example, one can choose $\hat{H}=\sigma_x/2$ and $\hat{h}=-\sigma_x/2$ (when $\Delta t\coloneqq t_1-t_0=\pi$, this Hamiltonian evolves the time eigenstates correctly:  $e^{-i \hat{h} \Delta t}\ket{0}=i \ket{1}$). Since now the clock is finite dimensional, Eq. \eqref{th} does not hold. Instead, the commutator is now $\left[\hat{t},\hat{h}\right]=-i \Delta t/2\,\sigma_y$.

A suitable Heisenberg state of system+clock is now $\ket{0}_{\mathfrak{S}}\left(\ket{0}_{\mathfrak{C}}+\ket{1}_{\mathfrak{C}}\right)/\sqrt(2)$. Under the unitary $\hat{\mathbb{U}}(t_0)\coloneqq e^{-i\hat{H}\otimes (\hat{t}-t_0\mathds{1}_{\mathfrak{C}})}$, the generators $\left(\sigma_x\otimes\mathds{1}_{\mathfrak{C}}, \sigma_z\otimes\mathds{1}_{\mathfrak{C}}\right)$ become 
\begin{align}
    &\left(\sigma_x\otimes\mathds{1}_{\mathfrak{C}},\right. \\
    &\left. \sigma_z \otimes \left[\ket{0}\bra{0} + \text{cos}\left(\Delta t\right)\ket{1}\bra{1}\right] +\text{sin}\left(\Delta t\right)\sigma_y\otimes\ket{1}\bra{1}\right).\nonumber 
\end{align}
This means that if the clock is in the state $\ket{0}_{\mathfrak{C}}$, the reduced relative generators of the system will be $\left(\sigma_x,\sigma_z \right)$; if the clock is in the state $\ket{1}_{\mathfrak{C}}$ the generators will be $\left(\sigma_x,\text{cos}\left(\Delta t\right)\sigma_z +\text{sin}\left(\Delta t\right)\sigma_y\right)$, as expected.

Finally, let me check that these results are consistent with the state of the system in the SP. To do so, let me write the density matrix of the system as 
\begin{equation}
    \rho_{\mathfrak{S}}(t)=\frac{1}{2}\left(\mathds{1}+\sum_i \lambda_i(t) \sigma_i\right),
    \label{rho_desc}
\end{equation}
with $\lambda_i(t) \coloneqq \text{Tr}\left[\rho_{\mathfrak{S}}(t)\sigma_i\right]=\text{Tr}\left[\rho_{\mathfrak{S}}(t_0)\sigma_i(\Delta t)\right]$, where $\left\{\sigma_i(\Delta t)\right\}_i$ are the Heisenberg-evolved Pauli matrices after a time $\Delta t$ and $\rho_{\mathfrak{S}}(t_0)=\ket{0}\bra{0}_{\mathfrak{S}}$. As I have shown above, at time $t_0$ the Pauli matrices are unchanged and thus, trivially, $\rho_{\mathfrak{S}}(t_0)=\ket{0}\bra{0}_{\mathfrak{S}}$. Using the Heisenberg-evolved generators above one finds at time $t_1$: $\left\{\lambda_i(t_1)\right\}_i=\left(0,- \text{sin}(\Delta t),\text{cos}(\Delta t)\right)$. Plugging these values into Eq. \eqref{rho_desc} one gets 
\begin{equation}
    \rho_{\mathfrak{S}}(t_1)=\frac{1}{2}
    \begin{pmatrix}
    1+\text{cos}(\Delta t) & i\,\text{sin}(\Delta t) \\    -i\,\text{sin}(\Delta t) & 1-\text{cos}(\Delta t)
    \end{pmatrix}.
\end{equation}
One would have obtained this density matrix by evolving the system's state in the SP.

\subsection{Qubit interacting with an ideal clock.}\label{ex_int_qubit}
Here, I will consider a qubit interacting with the clock. For simplicity, I will study the system in the SP, but the HP can be easily recovered using the results of Sec. \ref{HP}. Let me assume that the clock is ideal to guarantee enough degeneracy in the eigenvalues of the Hamiltonian of the universe. Let me assume a system-clock interaction as in Eq. \eqref{V_G}, with  $\hat{H}=\sigma_x/2$.
If the state of the qubit at time $t=0$ is $\ket{0}$, then the state of the universe with eigenvalue $\mathcal{E}$ is given by
\begin{equation}
    \kket{\Psi_\mathcal{E}}=\int \text{d}t \,e^{-i\frac{\hat{H}-\mathcal{E}}{1+\hat{H}/\Lambda}t}\ket{0}\ket{t}.
   \label{pure_int_qubit}
\end{equation}
The effective Hamiltonian of the system can be written as
\begin{equation}
    \hat{H}_{eff}(\mathcal{E})=\frac{\sigma_x/2-\mathcal{E}}{\mathds{1}+\frac{\sigma_x}{2\Lambda}}=\phi(\Lambda,\mathcal{E}) \mathds{1}+\omega(\Lambda,\mathcal{E}) \frac{\sigma_x}{2},
    \label{Heff_qubit}
\end{equation}
where
\begin{gather}
    \phi(\Lambda,\mathcal{E})=-\frac{1}{\Delta}\left(\mathcal{E}+\frac{1}{4\Lambda}\right),\label{phi}\\
    \omega(\Lambda,\mathcal{E}) = \frac{1}{\Delta}\left(1+\frac{\mathcal{E}}{\Lambda}\right)\label{omega},
\end{gather}
with $\Delta\coloneqq \text{det}(\mathds{1}+\frac{\sigma_x/2}{\Lambda})=1-\frac{1}{4\Lambda^2}$.
The Hamiltonian is well defined for $\Lambda\neq \pm 1/2$ and $\lambda\neq0$. For $\Lambda\rightarrow \pm 1/2$, $\phi$ and $\omega$ diverge. When $\Lambda =  1/2$ $\left(\Lambda = - 1/2\right)$, there is still a stationary state of the universe, namely $\ket{+}\ket{h=\mathcal{E}-1/4}$ $\left(\ket{-}\ket{h=\mathcal{E}+1/4}\right)$, but the system is stationary in the state $\ket{+}$ $\left(\ket{-}\right)$. When $\Lambda =0 $ the interaction term is not defined, but for  $\Lambda\rightarrow0$, $\hat{H}_{eff}\rightarrow0$ and thus the system is stationary (in the initial state $\ket{0}$).
Notice that when $\Lambda\rightarrow\infty$ (no interaction), $\hat{H}_{eff}\rightarrow\sigma_x/2$ for any value of $\mathcal{E}$, as expected.

\begin{figure}
	\centering	
	\includegraphics[width=8.6cm]{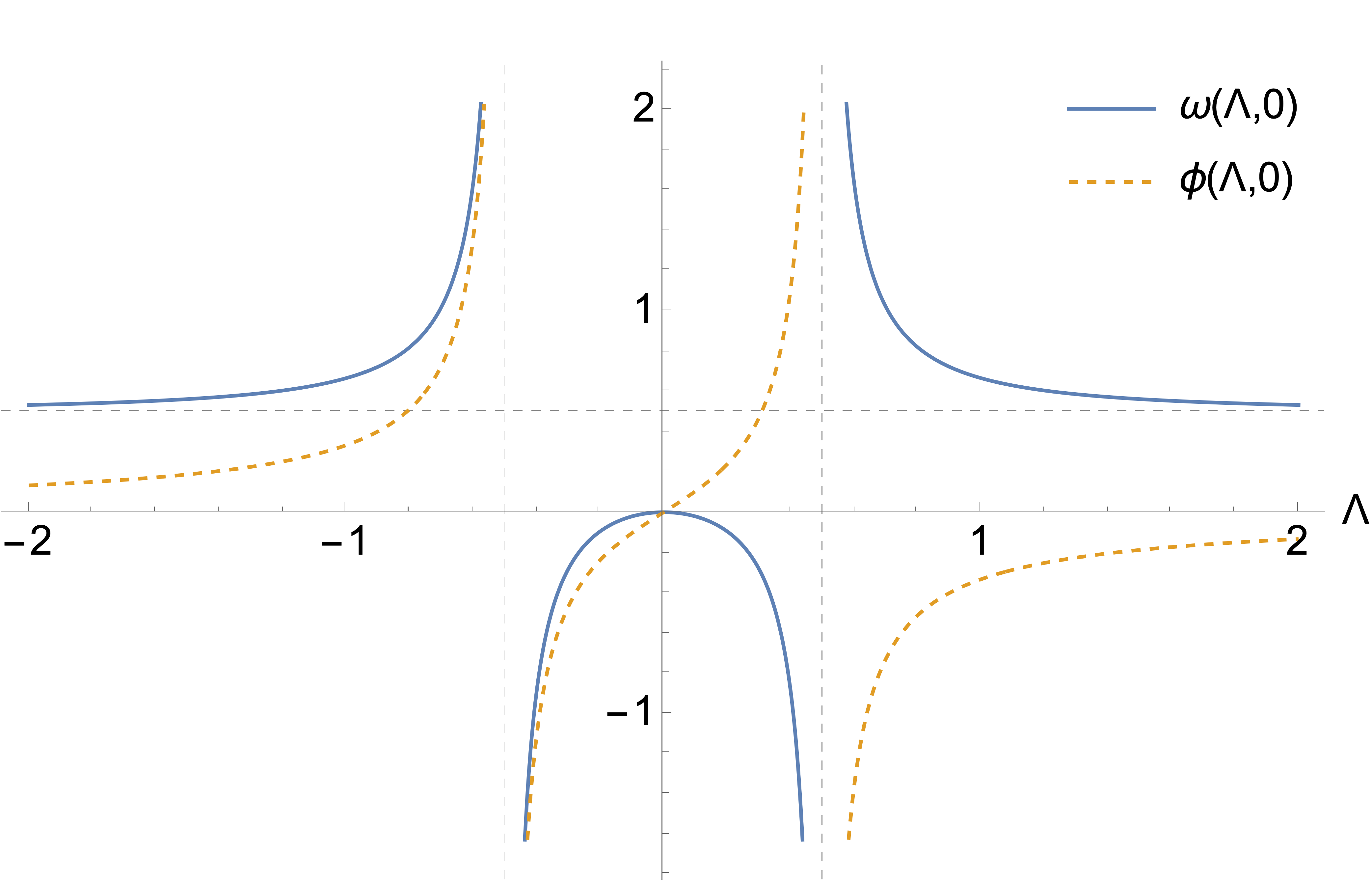}	
	\caption{Plot of the parameters governing the effective Hamiltonian of the two-level system discussed in Sec. \ref{ex_int_qubit}. $\omega$ and $\phi$ are plotted as functions of the interaction strength $\Lambda$ for $\mathcal{E}=0$ (in arbitrary units). $\phi$ contributes to an unobservable global phase while $\omega$ represents a time-scale factor for the system's evolution. Notice the inversion in the direction of the evolution for $-1/2<\Lambda<1/2$ and the divergences for $\Lambda=\pm 1/2$.}
	\label{parameters_nomass}
\end{figure}

The unitary evolution of the system is given by
\begin{equation}
    \hat{U}_{eff}(t) \coloneqq e^{-i\phi(\Lambda,\mathcal{E}) t}e^{-i\omega(\Lambda,\mathcal{E}) \sigma_x t/2}.
\end{equation}
The first part induces an unobservable phase shift while the second rotates the state of the qubit. The effect of the system-clock interaction is to change the speed at which the qubit evolves. The observable part of the effect is determined by $\omega$.

In Fig. \ref{parameters_nomass}, I have plotted the value of $\omega(\Lambda,0)$ and $\phi(\Lambda,0)$ for different values of $\Lambda$. For $\Lambda\rightarrow1/2$, the speed of the system's evolution diverges. Interestingly, for values of $\Lambda$ between $0$ and $1/2$ the system evolves back in time (the usual direction of time is defined by the non-interacting limit $\Lambda\rightarrow\infty$). For $\Lambda<0$, the situation is symmetrical for $\omega$ and antisymmetrical for $\phi$.
As can be seen in Eqs. \eqref{phi}-\eqref{omega}, for $\mathcal{E}\neq0$ there are additional zeroes for $\phi$ and $\omega$. When $\omega=0$ the system is stationary.

\subsection{Massive qubit interacting with an ideal clock.}\label{ex_massive_int_qubit}

Let me consider a system with mass $m$ and a two-level internal Hamiltonian $\hat{H}=mc^2+E_I\sigma_x/2$, where $E_I$ is the system's internal energy.
The effective Hamiltonian for this system under the system-clock interaction of Eq. \eqref{V_G} is still given by Eq. \eqref{Heff_qubit} but now with coefficients
\begin{gather}
    \phi_m(\Lambda,\mathcal{E})=\frac{mc^2}{\Delta}\left(1+\frac{mc^2-\mathcal{E}}{\Lambda} -\frac{\mathcal{E}+E_I^2/4\Lambda}{mc^2} \right),\label{phim}\\
    \omega_m(\Lambda,\mathcal{E}) = \frac{E_I}{\Delta}\left(1+\frac{\mathcal{E}}{\Lambda}\right)\label{omegam},
\end{gather}
with $\Delta=(1+\frac{mc^2}{\Lambda})^2-\frac{E_I^2}{4\Lambda^2}$. In the following, I will assume that the internal energy of the system is much smaller than its rest energy ($E_I\ll mc^2$).
\begin{figure}
	\centering	
	\includegraphics[width=8.6cm]{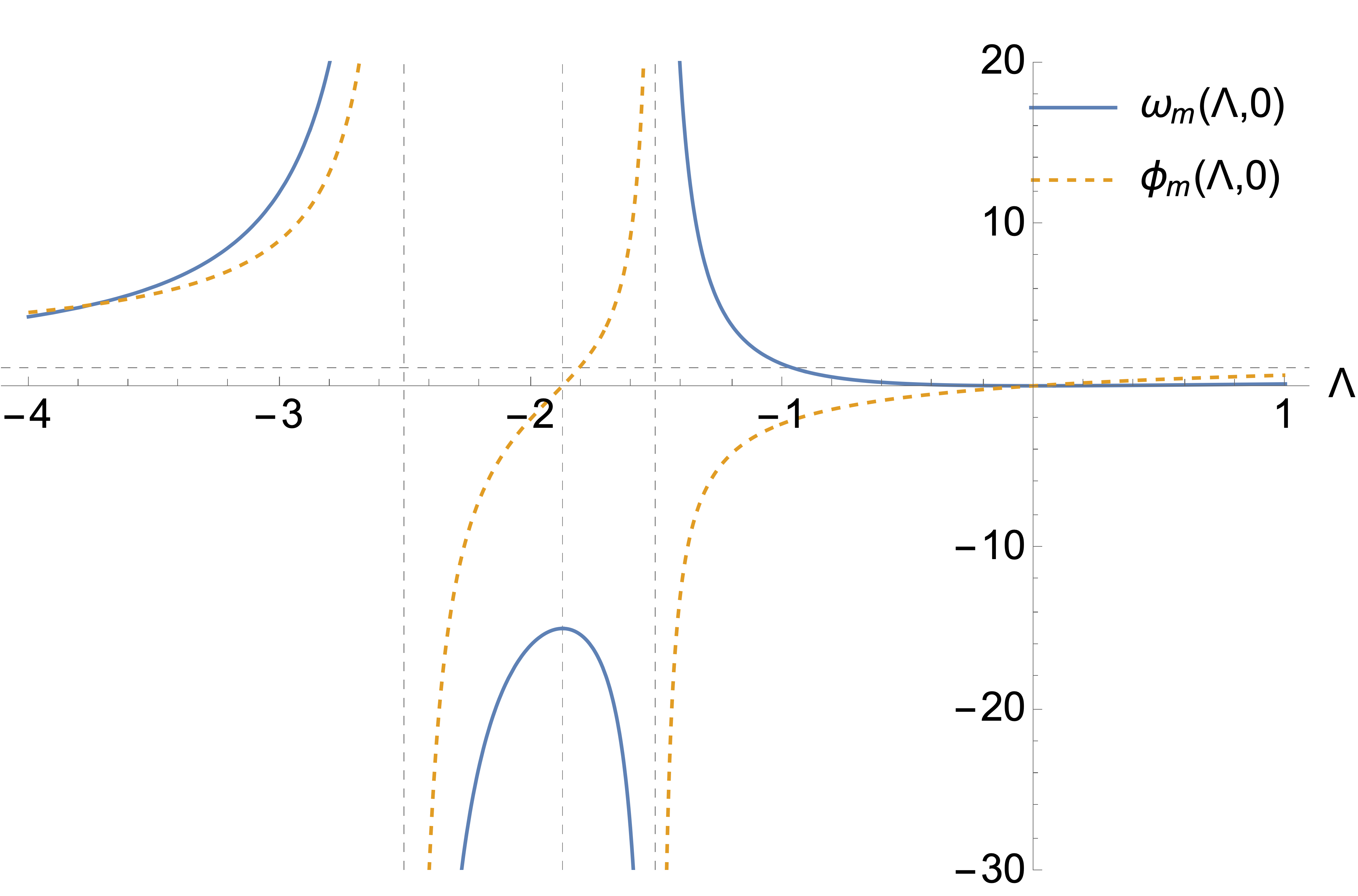}	
	\caption{Plot of the parameters governing the effective Hamiltonian of the massive two-level system discussed in Sec. \ref{ex_massive_int_qubit}. $\omega_m$ and $\phi_m$ are plotted as functions of the interaction strength $\Lambda$ for $\mathcal{E}=0$, $E_I=1$ and $mc^2=2$ (in arbitrary units). $\phi_m$ contributes to an unobservable global phase while $\omega_m$ represents a time-scale factor for the system's evolution. Notice the inversion in the direction of the evolution and the divergences.}
	\label{parameters_mass}
\end{figure}
The observable part of the evolution is still governed by $\omega_m$, which is plotted in Fig. \ref{parameters_mass}. The behaviour of $\omega_m$ is similar to the previous case, but now the divergences happen at $\Lambda=-mc^2\pm E_I/2$.
If one assumes a Newtonian system-clock interaction \cite{smith_quantizing_2019} $\Lambda=-dc^4/G$, with $d$ the clock-system distance, then the divergences happen at the critical values of $d$ 
\begin{equation}
    d_{\text{crit}}=\frac{R_S}{2}\mp \frac{E_I G}{2c^4},
\end{equation}
where $R_s=\frac{2Gm}{c^2}$ is the Schwarzschild radius of the system.
Between these two distances, the system evolves back in time.
Notice that if the system were a black hole and the system-clock coupling factor were $\Lambda=-dc^4/2G$, this time inversion would happen in a small region around the event horizon.
Similarly to the previous case, $\hat{H}_{eff}\rightarrow0$ for $\Lambda\rightarrow0$, and thus the system becomes stationary in this limit.

Finally, let me discuss what happens when the universe is in a mixed state $\hat{\mathcal{R}}$.
Since the spectrum of $\hat{\mathcal{H}}$ is continuous, there are infinite choices of $\hat{
\mathcal{R}}$ compatible with the system being initially in the state $\ket{0}\bra{0}$. For simplicity, I will choose $\hat{\mathcal{R}}$ to be an equal mixture of two eigensectors of $\hat{\mathcal{H}}$, one with eigenvalue $0$ and one with eigenvalue $\mathcal{E}$. The state of the system at time $t$ is thus given by Eq. \eqref{relative_mixed_int} 
\begin{align}
    \hat{\rho}_{\mathfrak{S}}(t)=\frac{1}{2}\left(  e^{-i \hat{H}_{eff}(0)t}\ket{0}\bra{0}e^{i \hat{H}_{eff}(0)t} + \right.\\ \nonumber
     \left.+ e^{-i \hat{H}_{eff}(\mathcal{E})t}\ket{0}\bra{0}e^{i \hat{H}_{eff}(\mathcal{E})t}\right).
\end{align}
Substituting the explicit expression for $\hat{H}_{eff}$ one gets
\begin{align}
    \hat{\rho}_{\mathfrak{S}}(t)=\frac{1}{2} e^{-i \frac{E_I t}{2 \Delta}\sigma_x }&\left(  \ket{0}\bra{0} + \vphantom{e^{-i \frac{E_I \mathcal{E} t}{2 \Delta \Lambda}\sigma_x}} \right.\\ \nonumber
   &  \left.+ e^{-i \frac{E_I \mathcal{E} t}{2 \Delta \Lambda}\sigma_x}\ket{0}\bra{0}e^{i\frac{E_I \mathcal{E} t}{2 \Delta \Lambda}\sigma_x}\right)e^{i \frac{E_I t }{2 \Delta}\sigma_x}.
\end{align}
The evolution of $ \hat{\rho}_{\mathfrak{S}}(t)$ is manifestly non-unitary. Interestingly, $ \hat{\rho}_{\mathfrak{S}}(t)$ reaches the maximally mixed state after a time
\begin{equation}
    \tau_D=\frac{\pi \hbar \Delta \Lambda}{E_I \mathcal{E}}=\frac{\pi \hbar \Lambda }{E_I \mathcal{E} } \left[\left(1+\frac{mc^2}{\Lambda}\right)^2 -\frac{E_I^2}{4\Lambda^2}\right],
\end{equation}
where I have reinstated $\hbar$.
$\tau_D$ has the meaning of a coherence time. 

Another meaningful quantity is the number of full rotations that the system can undergo before it reaches the maximally mixed state:
\begin{equation}
    \frac{\tau_D}{\tau}=\frac{\Lambda}{2\mathcal{E}},
\end{equation}
where ${\tau}=\frac{2\pi \hbar \Delta }{E_I}$ is the time it takes for a system to go back to the initial state in the case of no clock-system interactions. This quantity shows the two parameters that regulate the process. Strong system-clock coupling and big differences between the eigenvalues of $\hat{\mathcal{H}}$ lead to short ``relative'' coherence times for the system. 
Substituting $\Lambda=dc^4/G$ (the minus sign does not change the conclusions), one gets
\begin{equation}
\frac{\tau_D}{\tau}=\frac{dc^4}{2G\mathcal{E}}.
\end{equation} 
Interestingly, if $\mathcal{E}=mc^2$, $\tau_D/\tau=d/R_S$, where $R_S$ is the Schwarzschild radius of the system.

To grasp the magnitude of this effect, let me consider an atomic-scale system.  Assuming that the clock is gravitationally coupled to the system at a distance $d=\SI{e-10}{\m}$, ${\tau_D}/{\tau} \approx 10^{51}$ for $\mathcal{E}=\SI{10}{\eV}$, while ${\tau_D}/{\tau}\approx 10^{41}$ for $\mathcal{E}=m_p c^2$, where $m_p$ is the mass of the proton. Both these relative coherence times are very long.
These results show that even if the system's evolution is non-unitary, this effect may not be easy to detect for weak gravitational system-clock interactions.
Instead, the effect becomes significant for large values of $\mathcal{E}$ at scales where one would expect strong relativistic effects, e.g. for $\mathcal{E}=mc^2$ and $d=R_S$. Alternatively, the effect could increase with the number of different energy eigenvalues appearing in the state of the universe. I leave the investigation of this point open for future work.

\section{Conclusions.}
In this work, I have investigated some features and generalisations of the Page-Wootters  construction. First, I have shown how to formulate the PW construction in the Heisenberg picture. This is achieved with a formal unitary transformation that transfers the time dependence from the state of the universe to the observables. The same transformation applies to both pure and mixed states of the universe as long as there is no system-clock interaction.
In this picture, the dependence of the system's observables on the time operator of the clock is manifest, while the usual time evolution of quantum theory emerges in the relative observables. 

The Heisenberg picture formulation of the PW construction reveals some interesting aspects about the system-clock correlations. For instance, entanglement is not necessary for a non-trivial time evolution if one allows mixed states of the universe. One can encode the relevant system-clock correlations in a separable (mixed) state. In fact, for any stationary entangled state of the universe, it is possible to find a separable stationary mixed state that encodes the same time evolution.

In this work, I have also studied the role of some types of system-clock interactions in the PW construction. Under such interactions, different pure stationary states of the universe lead to different dynamics at the level of the system.
This is important because any realistic system must have some level of interaction between its subsystems. If the system under consideration is the whole universe, usually its state is assumed to be pure and with energy 0. However, this does not have to be the case for a generic closed system. Pure states of the universe with different energy eigenvalues lead to different effective Hamiltonians for the system, and mixed states of the universe can lead to a non-unitary evolution. In principle, it is possible to detect the latter effect and thus distinguish some mixed (stationary) states of the universe from the pure ones (contrary to \cite{page_evolution_1983}).
For instance, in the case of a two-level system interacting with an ideal clock, I showed that the non-unitarity of the system's dynamics can lead to the loss of coherence in a characteristic time that depends on the strength of the interaction. This effect is very small for weakly interacting systems, but it becomes important at scales where one would expect relativistic effects to arise. \\
Interestingly, at these scales, even pure states of the universe lead to unusual dynamics at the level of the system. For example, if the clock interacts gravitationally with a two-level massive system, there is a region of system-clock distances where the system evolves backwards in time. This happens when the system-clock distance is around half of the Schwarzschild radius of the system. I shall explore this effect thoroughly in a forthcoming paper.

The study of the PW construction with system-clock interactions shows another interesting aspect: the system can undergo a unitary evolution only if the state of the universe is entangled. In some sense, interactions restore the role of entanglement in the PW construction.
This could have interesting consequences for classical relational approaches to time \cite{foti_time_2021,vedral_classical_2022,anderson_problem_2017}. I leave the exploration of this point to future work.

The construction outlined here could be easily extended to include all the other generators of symmetry transformations \cite{singh_quantum_2021,hohn_internal_2022,de_la_hamette_perspective-neutral_2021}. It could also be applied to other related problems, such as indefinite causal order \cite{baumann_noncausal_2022}, the questions on clock synchronisation raised in \cite{kuypers_quantum_2022}, and time dilation \cite{smith_quantum_2020,paczos_quantum_2022,singh_emergence_2023}. 

Finally, it would be interesting to consider interactions such that $\hat{t}$ is not left invariant when moving from the Schr\"odinger picture to the Heisenberg picture. In this case, one has to modify the notion of partial trace to get the right relative state of the system in the Heisenberg picture. I leave this problem open for future investigations.

\section*{Acknowledgments.}
The author thanks Samuel Kuypers, Chiara Marletto, Nicetu Tibau Vidal, Vlatko Vedral, Aditya Iyer and Giuseppe Di Pietra for several fruitful discussions and comments. The author also thanks Ashmeet Singh and Oliver Friedrich for some stimulating conversations. This research was supported by the Foundation Blanceflor and the Fondazione ``Angelo Della Riccia''.

\appendix

\section{Properties of the SP-HP transformation for pure states of the universe.}

\subsection{Allowed transformations and gauge invariance.}\label{propUpure}
In section \ref{HPpure}, I have found that the unitary $\hat{\mathbb{U}}$ leads to the HP version of the PW construction. However, one may wonder if this unitary is the only possible transformation between the SP and the HP.
In other words, one has to motivate the choice of the unitary $\hat{\mathbb{U}}$ and of the Heisenberg state $\kket{\Psi_{\mathcal{E}}^{(\mathbb{H})}}$ such that $\hat{\mathbb{U}}\kket{\Psi_{\mathcal{E}}^{(\mathbb{H})}}=\kket{\Psi_{\mathcal{E}}}$.
Assuming that the transformation between the SP and the HP is unitary, any such choice is described by an additional unitary transformation $\hat{W}$: $\kket{\Psi_{\mathcal{E}}}=\hat{\mathbb{U}}\kket{\Psi_{\mathcal{E}}^{(\mathbb{H})}}=\hat{\mathbb{U}}\hat{W}\hat{W}^{\dagger}\kket{\Psi_{\mathcal{E}}^{(\mathbb{H})}}=\hat{U}'\kket{{\Psi'}_{\mathcal{E}}^{(\mathbb{H})}}$, where $\hat{U}'\coloneqq\hat{\mathbb{U}}\hat{W}$ and $\kket{{\Psi'}_{\mathcal{E}}^{(\mathbb{H})}}\coloneqq\hat{W}^{\dagger}\kket{\Psi_{\mathcal{E}}^{(\mathbb{H})}}$.
The choice of $\hat{U}'$ and $\kket{{\Psi'}_{\mathcal{E}}^{(\mathbb{H})}}$ is then determined by the requirements of Sec. \ref{HPpure}:
\begin{enumerate}
    \item $\hat{U}'^{\dagger}\left(\mathds{1}_\mathfrak{S}\otimes \hat{t}\right)\hat{U}' = \mathds{1}_\mathfrak{S}\otimes \hat{s}$, with $\hat{s}$ a self-adjoint operator of the clock, so that the partial inner product and partial trace of Eqs. \eqref{relative_state_S} and \eqref{relative_rho_S} are still meaningful. This condition is satisfied in all the cases considered in this work.

    \item The transformation between the SP and the HP transfers all the ``encoded time evolution'' from the state of the universe to the observables. In other words, $\kket{{\Psi'_{\mathcal{E}}}^{(\mathbb{H})}}$ is such that 
    \begin{equation}
        \prescript{}{\mathfrak{C}}{\brakket{\Bar{s}'}{{\Psi'_{\mathcal{E}}}^{(\mathbb{H})}}} = e^{i\phi(s,s')} \prescript{}{\mathfrak{C}}{\brakket{\Bar{s}}{{\Psi'_{\mathcal{E}}}^{(\mathbb{H})}}},
        \label{Hstate_condition}
        \end{equation}
        for all $s,s'$, where $\ket{\Bar{s}}$ ($\ket{\Bar{s}'}$) is an eigenstate of $\hat{s}$ with eigenvalue $s$ ($s'$) and $\phi(s,s')$ is a phase.
\end{enumerate}
Clearly, any local unitary $\hat{W}=\hat{W}_{\mathfrak{S}}\otimes\hat{W}_{\mathfrak{C}}$ is allowed since $\hat{W}$ leaves $\mathds{1}_\mathfrak{S}\otimes \hat{t}$ of the form $\mathds{1}_\mathfrak{S}\otimes \hat{s}$ and $\prescript{}{\mathfrak{C}}{\brakket{\Bar{s}'}{{\Psi'_{\mathcal{E}}}^{(\mathbb{H})}}} \propto \hat{W}_{\mathfrak{S}}  \prescript{}{\mathfrak{C}}{\brakket{s'}{{\Psi_{\mathcal{E}}}^{(\mathbb{H})}}} \propto \hat{W}_{\mathfrak{S}}  \prescript{}{\mathfrak{C}}{\brakket{s}{{\Psi_{\mathcal{E}}}^{(\mathbb{H})}}} \propto \prescript{}{\mathfrak{C}}{\brakket{\Bar{s}}{{\Psi'_{\mathcal{E}}}^{(\mathbb{H})}}}$ $\forall\,s,s'$ ($\ket{s}$ is an eigenstate of $\hat{t}$ with eigenvalue $s$).
Since in ordinary quantum theory it is possible to change the basis with impunity, this is a desirable feature of the construction. However, these local changes of basis are not of much interest for the current purposes, so I will only consider local unitaries on the clock's Hilbert space that leave $\hat{t}$ invariant and do not change the basis of the system's Hilbert space.

In appendix \ref{nonlocal_pure}, I show that any non-local unitary $\hat{W}$ is not allowed.
The class of unitaries and Heisenberg states that I will consider is thus:
\begin{gather}
    \hat{\mathbb{U}}_{\phi} \coloneqq \hat{\mathbb{U}}e^{-i\mathds{1}_{\mathfrak{S}}\otimes\phi(\hat{t})}, \\
    \kket{\Psi_{\mathcal{E},\phi}^{(\mathbb{H})}} \coloneqq e^{i\mathds{1}_{\mathfrak{S}}\otimes\phi(\hat{t})}\ket{\psi(0)}_{\mathfrak{S}}\ket{\text{TL}_{\mathcal{E}}}_{\mathfrak{C}},
\end{gather}
with $\phi(\hat{t})$ a self-adjoint function of $\hat{t}$ only. Notice that $e^{i\mathds{1}_{\mathfrak{S}}\otimes\phi(\hat{t})}$ does not change either the observables or the relative observables of the system.
The simplest choice is $\hat{\mathbb{U}}$ and $\kket{\Psi_{\mathcal{E}}^{(\mathbb{H})}}$. For this choice, the stationarity constraint on the state of the universe becomes:
\begin{equation}
    \left(\mathds{1}_{\mathfrak{S}}\otimes\hat{h}\right) \kket{\Psi_{\mathcal{E}}^{(\mathbb{H})}}=\mathcal{E}\kket{\Psi_{\mathcal{E}}^{(\mathbb{H})}}.
    \label{stat_pure_HP}
\end{equation}
%In Appendix \ref{appB}, I also show that $\hat{\mathbb{U}}$ is the only unitary such that $\hat{\mathbb{U}}\ket{\psi(0)}_{\mathfrak{S}}\ket{\text{TL}_{\mathcal{E}}}_{\mathfrak{C}}=\kket{\Psi_\mathcal{E}}$ for any energy $\mathcal{E}$ and for any state $\ket{\psi(0)}_{\mathfrak{S}}$, where $\ket{\text{TL}_{\mathcal{E}}}_{\mathfrak{C}}\coloneqq \int \text{d}t\, e^{i\mathcal{E} t}\ket{t}_{\mathfrak{C}}=\sqrt{2\pi}\ket{h=\mathcal{E}}_{\mathfrak{C}}$.

Notice that, since $\hat{W}$ is a local unitary, any Heisenberg state must be \underline{separable} with respect to clock and system (see Appendix \ref{stat_sep} for another proof of this).
\vspace{0.5cm}

Let me now discuss some additional features of the construction.
First, let me notice that there is what one may call a gauge freedom. In the HP, the unitaries $\hat{\mathcal{U}}(\theta)\coloneqq \mathds{1}_{\mathfrak{S}} \otimes e^{-i\hat{h}\theta}$ are gauge transformations, since $\hat{\mathcal{U}}(\theta)\kket{\Psi^{(\mathbb{H})}_{\mathcal{E},0}}=e^{-i\mathcal{E} \theta}\kket{\Psi^{(\mathbb{H})}_{\mathcal{E},0}} \hspace{0.1cm} \forall \,\theta, \mathcal{E} \in \mathds{R}$.
Under $\hat{\mathcal{U}}(\theta)$, the time operator $\hat{t}$ undergoes a time translation: $\hat{\mathcal{U}}^{\dagger}(\theta)\left(\mathds{1}_{\mathfrak{S}}\otimes\hat{t}\right)\hat{\mathcal{U}}(\theta)=\mathds{1}_{\mathfrak{S}}\otimes\left(\hat{t}-\theta \hat{\mathds{1}}_{\mathfrak{C}}\right) $,
and all the observables related to $\hat{O}^{(\mathbb{H})}$ by $\hat{\mathcal{U}}(\theta)$
\begin{equation}
\hat{\mathcal{U}}^{\dagger}(\theta)\hat{O}^{(\mathbb{H})}\hat{\mathcal{U}}(\theta)=e^{i\hat{H}\otimes\left(\hat{t}- \theta \hat{\mathds{1}}_{\mathfrak{C}} \right)}\hat{O} e^{-i\hat{H}\otimes\left(\hat{t}- \theta \hat{\mathds{1}}_{\mathfrak{C}} \right)},
\end{equation}
belong to the same equivalence class. The observables found in, for example, \cite{hohn_trinity_2021} are the so-called ``group average'' of the observables in these equivalence classes \cite{bartlett_reference_2007,hohn_trinity_2021}. Notice that the group average of the system's observables in these equivalence classes results in observables that commute with the Hamiltonian of the universe.

For a general closed system, these new observables describe a different physical situation even if all the expectations values are the same \cite{deutsch_vindication_2012}. The difference lies in the different dynamics described by the old and the new observables, and it can be detected by ``starting'' with another Heisenberg state that is not an eigenstate of $\hat{\mathcal{U}}(\theta)$. However, the PW construction does not allow states that do not satisfy the stationarity condition, so this difference is impossible to detect.

This reasoning may not apply to other transformations that have $\kket{\Psi^{(\mathbb{H})}}$ as an eigenstate (for instance, a local transformation on the system). Still, it is unclear what it means to start with a different Heisenberg state of the universe and, thus, whether it is possible to detect the difference between the new and the old observables, even in principle. I leave the questions regarding this possibility open for future discussion.

\subsection{Condition on the unitary transformations of $\hat{\mathbb{U}}$.}\label{nonlocal_pure}
Given that $\kket{\Psi}=\hat{\mathbb{U}}\kket{\Psi^{(\mathbb{H})}}=\hat{\mathbb{U}}\hat{W}\hat{W}^{\dagger}\kket{\Psi^{(\mathbb{H})}}$, let me assume that $\hat{W}$ leaves $\hat{t}$ invariant. This means that $\hat{W}$ must be a function of the system's operators and $\hat{t}$ only. Thus, $\hat{W}=\int \text{d}t\,\hat{A}(t)\otimes\ket{t}\bra{t}$, with $\hat{A}$ unitary.
The condition on $\kket{\Psi'^{(\mathbb{H})}}$
\begin{equation}
    \brakket{t'}{\Psi'^{(\mathbb{H})}}=e^{i\phi(t,t')}\brakket{t}{\Psi'^{(\mathbb{H})}} \hspace{0.3cm} \forall \, t,t',
    \label{appB1}
\end{equation}
implies that 
\begin{equation}
    \hat{A}^{\dagger}(t')\brakket{t}{\Psi^{(\mathbb{H})}}=e^{i\phi(t,t')}\hat{A}^{\dagger}(t)\brakket{t}{\Psi^{(\mathbb{H})}} \hspace{0.3cm} \forall \, t,t',
    \label{appB2}
\end{equation}
where I have used that fact that $\brakket{t}{\Psi^{(\mathbb{H})}}=\brakket{t'}{\Psi^{(\mathbb{H})}}\hspace{0.1cm}\forall\,t,t'$.
If one also requires that $\hat{W}^{\dagger}\kket{\Psi^{(\mathbb{H})}}$ is such that Eq. \eqref{appB1} holds for any state initial state $\ket{\psi}$ of the system, then 
\begin{equation}
    \hat{A}^{\dagger}(t')\ket{\psi}=e^{i\phi(t,t')}\hat{A}^{\dagger}(t)\ket{\psi}\hspace{0.3cm} \forall \, t,t'.
    \label{appB3}
\end{equation}
and for any $\ket{\psi}$. Therefore, $\hat{A}^{\dagger}(t')=e^{i\phi(t,t')}\hat{A}^{\dagger}(t)\hspace{0.1cm}\forall\,t,t'$ and thus $\hat{W}=\hat{A}(0)\otimes\int \text{d}t\,e^{-i\phi(0,t)}\ket{t}\bra{t}$, which is a local unitary.

What if $\hat{W}^{\dagger}\left(\mathds{1}_\mathfrak{S}\otimes \hat{t}\right)\hat{W}=\mathds{1}_\mathfrak{S}\otimes \hat{s}$?
In this case, it is always possible to find a local unitary of the clock $\hat{V}$ such that $\hat{V}^{\dagger}\left(\mathds{1}_\mathfrak{S}\otimes \hat{s}\right)\hat{V}=\mathds{1}_\mathfrak{S}\otimes \hat{t}$. This means that 
\begin{equation}
    \hat{W}^{\dagger}\hat{V}^{\dagger}\left(\mathds{1}_\mathfrak{S}\otimes \hat{s}\right)\hat{V}\hat{W}=\mathds{1}_\mathfrak{S}\otimes \hat{s},
    \label{appB4}
\end{equation} 
and thus the unitary $\hat{V}\hat{W}$ is a function of the observables of the system and $\hat{s}$ only. Moreover, $\kket{\Psi'^{(\mathbb{H})}}=\hat{W}^{\dagger}\hat{V}^{\dagger}\hat{V}\kket{\Psi^{(\mathbb{H})}}=\hat{W}^{\dagger}\hat{V}^{\dagger}\kket{{\Psi''}^{(\mathbb{H})}}=\left( \hat{V}\hat{W}\right)^{\dagger}\kket{{\Psi''}^{(\mathbb{H})}}$, where $\kket{{\Psi''}^{(\mathbb{H})}}\coloneqq\hat{V}\kket{\Psi^{(\mathbb{H})}}$ is a state for which \eqref{Hstate_condition} holds (since $\hat{V}$ is a local unitary). 
Therefore, one can apply the reasoning above to the unitary $\hat{V}\hat{W}$, with $\kket{{\Psi''}^{(\mathbb{H})}}$ instead of $\kket{{\Psi}^{(\mathbb{H})}}$ in Eq. \eqref{appB2}. Thus, $\hat{V}\hat{W}$ must be a local unitary and (since $\hat{V}$ is a local unitary) $\hat{W}$ must be a local unitary too.

\subsection{No encoded time evolution and separability.}\label{stat_sep}
The pure Heisenberg state of the universe must be such that $\prescript{}{\mathfrak{C}}{\brakket{s'}{\Psi'^{(\mathbb{H})}}} = e^{i\phi(s,s')} \prescript{}{\mathfrak{C}}{\brakket{s}{\Psi'^{(\mathbb{H})}}} \hspace{0.1cm} \forall\, s,s'$, where $\ket{s}$ and $\ket{s'}$ are eigenstates of the clock-observable $\hat{s}$ and $\phi(s,s')$ is a phase. This implies that 
\begin{align}
\kket{\Psi'^{(\mathbb{H})}}=&\int ds\,\ket{s}_{\mathfrak{C}}\bra{s}\cdot\kket{\Psi'^{(\mathbb{H})}}=\nonumber\\
=&\brakket{0}{\Psi'^{(\mathbb{H})}} \left(\int ds\,e^{i\phi(s,0)}\ket{s}_{\mathfrak{C}}\right)=\nonumber\\
=&\ket{\psi'(0)}_{\mathfrak{S}}\left(\int ds\,e^{i\phi(s,0)}\ket{s}_{\mathfrak{C}}\right) ,
\label{appA_suff}
\end{align}
which is a separable state of system and clock.

Vice versa, if $\kket{\Psi'^{(\mathbb{H})}}$ is separable, that is $\kket{\Psi'^{(\mathbb{H})}}=\ket{X}_{\mathfrak{C}}\ket{Y}_{\mathfrak{S}}$, then 
\begin{align}
    \brakket{s'}{\Psi'^{(\mathbb{H})}}=&\braket{s'}{X} \ket{Y}=e^{i\phi(s,s')}\braket{s}{X} \ket{Y}= \nonumber\\
    =&e^{i\phi(s,s')}\brakket{s}{\Psi'^{(\mathbb{H})}} \hspace{0.1cm} \forall\, s,s',
    \label{appA_nec}
\end{align}
where I have used the fact that $\kket{\Psi'^{(\mathbb{H})}}$ is normalised as in Eq. \eqref{initial_state} (so that $\braket{Y}_{\mathfrak{S}}=1$ and $\braket{s}{X}_{\mathfrak{C}}$ is just a phase for any $s$).
Let me notice that the case $\brakket{s}{\Psi'^{(\mathbb{H})}}=0$ is not allowed since this would imply that $\brakket{t}{\Psi}=0$ for some $\ket{t}$ eigenstate of $\hat{t}$ and thus (applying the stationarity condition of Eq. \eqref{stationarity}) $\brakket{t'}{\Psi}=0\hspace{0.1cm}\forall\,t'$. This implies that $\kket{\Psi}=0$, which is not an allowed state of the universe.

\section{Properties of the SP-HP transformation for mixed states of the universe.}

\subsection{Allowed transformations.}\label{propUmixed}
In section \ref{HP_mixed}, I have found that the unitary $\hat{\mathbb{U}}$ leads to the HP version of the PW construction for mixed states of the universe. The associated Heisenberg states are given by Eq. \eqref{mixedHstate}. Is this transformation between the two pictures unique?

Similarly to the conditions discussed in Appendix \ref{propUpure}, the Heisenberg state of the universe should not encode any time evolution, i.e. it should be such that Eq. \eqref{no_evo_mixed} holds. Since the universe I am considering is a closed system, every other possible Heisenberg state $\hat{\mathcal{R}}'^{(\mathbb{H})}$ should be linked to $\hat{\mathcal{R}}^{(\mathbb{H})}$ via a unitary transformation $\hat{\mathcal{R}}'^{(\mathbb{H})}=\hat{W}^{\dagger}\hat{\mathcal{R}}^{(\mathbb{H})}\hat{W}$\footnote{However, notice that $ \hat{\mathcal{R}}^{(\mathbb{H})}=\hat{\rho}_{\mathfrak{S}}(0) \otimes \ket{\text{TL}_{\mathcal{E}}}\bra{\text{TL}_{\mathcal{E}}}$ and $ \hat{\mathcal{R}}^{(\mathbb{H})}=\hat{\rho}_{\mathfrak{S}}(0) \otimes \mathds{1}_{\mathfrak{C}}$ are two Heisenberg states described by Eq. \eqref{mixedHstate} but they are not related by a unitary transformation.}.

Any local unitary $\hat{W}=\hat{W}_{\mathfrak{S}}\otimes\hat{W}_{\mathfrak{C}}$ is allowed since $\hat{W}$ leaves $\mathds{1}_\mathfrak{S}\otimes \hat{t}$ of the form $\mathds{1}_\mathfrak{S}\otimes \hat{s}$ and $\underset{\mathfrak{C}}{\text{Tr}}\left[\hat{\mathcal{R}}'^{(\mathbb{H})} \hat{\Pi}_s(\hat{s})\right]= \hat{W}_{\mathfrak{S}}^{\dagger}\underset{\mathfrak{C}}{\text{Tr}}\left[\hat{\mathcal{R}}^{(\mathbb{H})} \hat{\Pi}_s(\hat{t})\right] \hat{W}_{\mathfrak{S}}=  \hat{W}_{\mathfrak{S}}^{\dagger}\underset{\mathfrak{C}}{\text{Tr}}\left[\hat{\mathcal{R}}^{(\mathbb{H})} \hat{\Pi}_{s'}(\hat{t})\right] \hat{W}_{\mathfrak{S}}=\underset{\mathfrak{C}}{\text{Tr}}\left[\hat{\mathcal{R}}'^{(\mathbb{H})} \hat{\Pi}_{s'}(\hat{s})\right]$
    $\forall \, s,s'$.
    
In Appendix \ref{nonlocal_mixed}, I show that non-local unitaries are not allowed. 
Therefore, restricting to transformations that leave $\hat{t}$ invariant and do not change the basis of the system's Hilbert space, the  class of unitaries and Heisenberg states that I will consider is:
\begin{gather}
    \hat{\mathbb{U}}_{\phi} \coloneqq \hat{\mathbb{U}}e^{-i\mathds{1}_{\mathfrak{S}}\otimes\phi(\hat{t})}, \\
    \hat{\mathcal{R}}_{\phi}^{(\mathbb{H})} \coloneqq e^{i\mathds{1}_{\mathfrak{S}}\otimes\phi(\hat{t})}\hat{\mathcal{R}}^{(\mathbb{H})}e^{-i\mathds{1}_{\mathfrak{S}}\otimes\phi(\hat{t})},
\end{gather}
with $\phi(\hat{t})$ a self-adjoint function of $\hat{t}$ only.
The simplest choice is represented by $\hat{\mathbb{U}}$ and $\hat{\mathcal{R}}^{(\mathbb{H})}$.

\subsection{Condition on the unitary transformations of $\hat{\mathbb{U}}$.}\label{nonlocal_mixed}
Given that $\hat{\mathcal{R}}'^{(\mathbb{H})}=\hat{W}^{\dagger}\hat{\mathcal{R}}^{(\mathbb{H})}\hat{W}$, let me assume that $\mathds{1}_\mathfrak{S}\otimes \hat{t}$ is invariant under $\hat{W}$. This means that $\hat{W}$ must be a function of the system's operators and $\hat{t}$ only. Thus, $\hat{W}=\int \text{d}t\,\hat{A}(t)\otimes\ket{t}\bra{t}$, with $\hat{A}$ unitary.
Under this assumption, I will show that the unitary $\hat{W}$ must be a local unitary.

Let me start from the fact that  $\hat{\mathcal{R}}'^{(\mathbb{H})}$ must encode no time evolution, that is
\begin{equation}
    \underset{\mathfrak{C}}{\text{Tr}}\left[\hat{W}^{\dagger}\hat{\mathcal{R}}^{(\mathbb{H})}\hat{W} \hat{\Pi}_t(\hat{t})\right]= \underset{\mathfrak{C}}{\text{Tr}}\left[\hat{W}^{\dagger}\hat{\mathcal{R}}^{(\mathbb{H})}\hat{W} \hat{\Pi}_{t'}(\hat{t})\right]
    \hspace{0.3cm} \forall \, t,t'.
    \label{appC1}
\end{equation}
and thus
\begin{align}
   &\hat{A}^{\dagger}(t) \underset{\mathfrak{C}}{\text{Tr}}\left[\hat{\mathcal{R}}^{(\mathbb{H})} \hat{\Pi}_t(\hat{t})\right]\hat{A}(t)= \hat{A}^{\dagger}(t')\underset{\mathfrak{C}}{\text{Tr}}\left[\hat{\mathcal{R}}^{(\mathbb{H})} \hat{\Pi}_{t'}(\hat{t})\right]\hat{A}(t')=\nonumber\\
   &=\hat{A}^{\dagger}(t')\underset{\mathfrak{C}}{\text{Tr}}\left[\hat{\mathcal{R}}^{(\mathbb{H})} \hat{\Pi}_{t}(\hat{t})\right]\hat{A}(t')
    \hspace{0.3cm} \forall \, t,t'.
    \label{appC2}
\end{align}
where the second equality is due to Eq. \eqref{no_evo_mixed}.
I will further require that $\hat{W}^{\dagger}\hat{\mathcal{R}}^{(\mathbb{H})}\hat{W}$ must be such that Eq. \eqref{no_evo_mixed} holds for any choice of $\hat{\mathcal{R}}^{(\mathbb{H})}$ given by Eq. \eqref{mixedHstate}.
Thus, let me choose $\hat{\mathcal{R}}^{(\mathbb{H})}_{\ket{\psi}
,\mathcal{E}} =\ket{\psi}\bra{\psi} \otimes \ket{\text{TL}_{\mathcal{E}}}\bra{\text{TL}_{\mathcal{E}}}$ for any $\ket{\psi}$ and ${\mathcal{E}}$.
I can then rewrite Eq. \eqref{appC2} as 
\begin{align}
  \hat{A}^{\dagger}(t) 
 \ket{\psi}\bra{\psi}  \hat{A}(t)= \hat{A}^{\dagger}(t') 
 \ket{\psi}\bra{\psi}  \hat{A}(t')
    \hspace{0.3cm} \forall \, t,t',
    \label{no_evo_mixed3}
\end{align}
and for any state of the system $\ket{\psi}$.
This implies that  $\hat{A}(t')=e^{i\phi(t,t')}\hat{A}(t)$ and thus $\hat{W}=\hat{A}(0) \otimes \int \text{d}t\, e^{i\phi(t,0)}\ket{t}\bra{t}$ which is a local unitary. 
The demonstration for the case when $\mathds{1}_\mathfrak{S}\otimes \hat{t}$ is not invariant under $\hat{W}$ (but is still of the form $\mathds{1}_\mathfrak{S}\otimes \hat{s}$)  is similar to that in Appendix \ref{nonlocal_pure}.

\bibliography{main}

\end{document}